\documentclass{aastex631}

\usepackage{amsmath}

\newcommand\lya{$\mathrm{Lyman}\,\alpha$}

\newcommand{\mearth}{$\rm M_{\earth}$}

\newcommand{\Teff}{\mbox{$T_{\mathrm{eff}}$}}

\newcommand{\pid}{16701}

\begin{document}

\title{The MUSCLES Extension for Atmospheric Transmission Spectroscopy: Spectral energy distributions for 20 exoplanet host stars that JWST observed in Cycle 1}

\correspondingauthor{David J. Wilson}
\email{djwilson394\@gmail.com}

\author[0000-0001-9667-9449]{David J. Wilson}
\affil{Laboratory for Atmospheric and Space Physics, University of Colorado, 600 UCB, Boulder, CO 80309}

\author[0000-0002-1176-3391]{Allison Youngblood}
\affil{NASA Goddard Space Flight Center, Greenbelt, MD 20771}

\author[0000-0002-7119-2543]{Girish M. Duvvuri}
\affil{Department of Physics and Astronomy, Vanderbilt University, Nashville, TN 37235, USA}

\author[0000-0002-1002-3674]{Kevin France}
\affil{Department of Astrophysical and Planetary Sciences, University of Colorado, Boulder, CO 80309, USA}
\affil{Laboratory for Atmospheric and Space Physics, University of Colorado, 600 UCB, Boulder, CO 80309}

\author[0000-0002-5094-2245]{P.\ Christian Schneider}
\affil{Hamburger Sternwarte, Gojenbergsweg 112, 21029 Hamburg }

\author[0000-0003-2631-3905]{Alexander Brown}
\affil{Center for Astrophysics and Space Astronomy, University of Colorado, 389 UCB, Boulder, CO 80309}

\author[0009-0006-5319-8649]{Isabella Longo}
\affil{Laboratory for Atmospheric and Space Physics, University of Colorado, 600 UCB, Boulder, CO 80309}
\affil{Department of Electrical and Computer Engineering, University of Texas, Austin, TX 78712, USA}

\author[0000-0001-8499-2892]{Cynthia S. Froning}
\affiliation{Southwest Research Institute, 6220 Culebra Road, San Antonio, TX 78238, USA}

\author[0000-0003-4733-6532]{Jacob L.\ Bean}
\affil{Department of Astronomy and Astrophysics, University of Chicago, Chicago, IL 60637, USA}

\author[0000-0002-4489-0135]{J. Sebastian Pineda}
\affil{Laboratory for Atmospheric and Space Physics, University of Colorado, 600 UCB, Boulder, CO 80309}

\author[0000-0002-1337-9051]{Eliza M.-R. Kempton}
\affiliation{Department of Astronomy \& Astrophysics, University of Chicago, Chicago, IL 60637, USA}
\affiliation{Department of Astronomy, University of Maryland, College Park, MD 20742, USA}

\author[0000-0002-0747-8862]{Yamila Miguel}
\affil{Leiden Observatory, Einsteinweg 55, 2333 CC Leiden}
\affil{SRON Netherlands Institute for Space Research, Niels Bohrweg 4, 2333 CA Leiden, The Netherlands}

\author[0000-0001-5442-1300]{Thomas M.\ Evans-Soma}
\affil{School of Information and Physical Sciences, University of Newcastle, Callaghan, NSW, Australia}

\author[0000-0002-3321-4924]{Zachory Berta-Thompson}
\affil{Department of Astrophysical and Planetary Sciences, University of Colorado, Boulder, CO 80309, USA}



\begin{abstract}
Correctly interpreting JWST spectra of close-in exoplanets requires a measurement of the X-ray and ultraviolet light that the planets receive from their host stars. Here we provide spectral energy distributions (SEDs) covering the range $\approx5-1\times10^7$\,\AA\ for 20 transiting exoplanet host stars observed in JWST Cycle 1. The SEDs are constructed out of new and archival Hubble Space Telescope, Chandra X-ray Observatory and/or XMM-Newton data combined with spectra from models or stars with similar properties (proxies) filling in unobserved gaps. We have also constructed SEDs of likely Habitable Worlds Observatory targets $\kappa^1$\,Ceti, $\tau$\,Ceti, $\epsilon$\,Indi and 70 Oph\,B for use as proxies. We find that the JWST target planets almost all experience much stronger ultraviolet fluxes than the Earth, especially in the extreme ultraviolet, even for planets with similar overall instellation. Strong ongoing or past atmospheric escape is possible for a majority of these planets. We also assess the now considerable sample of panchromatic stellar SEDs and its applicability for current JWST observations and beyond. 

\end{abstract}



\section{Introduction} \label{sec:intro}
The launch of JWST has enabled observations of transiting exoplanet atmospheres with unprecedented detail and sensitivity \citep{kemptonetal24-1,espinoza+perrin25-1,kreidburg+stevenson25-1}. Atmosphere studies of close-in exoplanets rely on detailed characterization of their host stars, including their ultraviolet and X-ray emissions. Among the very first results from JWST was the detection of SO$_2$ in the atmosphere of the hot Jupiter WASP-39\,b \citep{tsaietal23-1,powelletal24-1}, which is produced by photochemistry driven by near- and far-ultraviolet (NUV/FUV) photons. For smaller terrestrial exoplanets, non-detections of thick atmospheres \citep[e.g.,][]{kreidbergetal19-1,greenetal23-1,ziebaetal23-1,zhangetal24-1} may be explained by atmospheric escape due to the high X-ray and extreme ultraviolet (EUV) emissions of their host stars.

Although X-ray and ultraviolet stellar astronomy has been carried out for several decades with a focus on understanding the physics of stellar atmospheres and activity \citep[e.g.][]{schmittetal95-1, walkowiczetal08-1, ostenetal16-1}, the last decade has seen a renewed interest in detailed high-energy stellar observations driven by the desire to better understand exoplanet atmospheres. Work on this topic has included large programs such as (Mega-)~MUSCLES \citep{franceetal16-1, loydetal16-1, wilsonetal25-1} and HAZMAT/KAT \citep{shkolniketal14-1, richey-yowelletal22-2}, along with many smaller projects \citep[e.g][]{guinanetal16-1, pinedaetal21-2, diamond-loweetal22-1, feinsteinetal22-1, engle24-1} These programs have considerably improved our understanding of (in particular) M and K dwarf stars as a function of both spectral type and age, and have provided a wide range of publicly available High Level Science Products (HLSPs). The HLSPs typically include full spectral energy distributions (SEDs) covering the X-ray through the ultraviolet to the optical and infrared, of which there are now several dozen available from the Mikulski Archive for Space Telescopes (MAST). 

Despite this archive, new observations are still required to provide SEDs for JWST host stars. Previous surveys have focused mostly on M\,dwarf stars, whilst the JWST observations span a wider range of spectral types so suitable existing observations and/or proxy stars may not have been available. Furthermore, it has been consistently found that even stars with similar characteristics can have different high-energy spectra \citep{tealetal22-1, wilsonetal25-1}, and scaling relationships between optical lines and ultraviolet lines and fluxes have considerable scatter \citep{youngbloodetal17-1, melbourneetal20-1}. In the X-ray there is a well-established relationship between rotation period and X-ray luminosity \citep[e.g][]{wrightetal18-1}, but again with scatter of roughly an order of magnitude. Even when observing a given star, the ultraviolet \citep{kamgaretal24-1} and X-ray \citep{ayres25-1} fluxes can vary considerable over a years-long stellar cycle. For all of these reasons, dedicated observations of each host star, close in time to the relevant exoplanet observation, remain the ideal. To this end, \citet{behretal23-1} produced the MUSCLES Extension for Atmospheric Transmission Spectroscopy (MEATS), providing SEDs for the host stars of JWST GTO and ERS exoplanet transmission spectroscopy targets. Here we present Mega-MEATS, expanding the sample with SEDs for 20 JWST GO Cycle 1 transiting exoplanet targets for which no or insufficient ultraviolet spectroscopy was available, based on Hubble Space Telescope (HST), XMM-Newton (XMM) and Chandra X-ray Observatory (Chandra) observations.

The SEDs will be made available along with this paper on the MUSCLES High Level Science Product page\citep{muscleshlsp}\footnote{\url{https://archive.stsci.edu/prepds/muscles/}}, with the same formats as described for Mega-MUSCLES in \citet{wilsonetal25-1}.  In this paper we describe the observations and models used to construct the SEDs (Section \ref{sec:sed}), with detailed notes on each star where necessary (Appendix \ref{sec:notes}). In Section \ref{sec:disc} we discuss the impacts of the high energy radiation experienced by the JWST planets and compare it to the Sun and Earth, and assess the state-of-the-art of available exoplanet host star SEDs. We also provide SEDs for four notable, nearby stars that are used here as proxies but which are also potential targets for the Habitable Worlds Observatory (HWO) (Appendix \ref{sec:moreseds}).

\begin{figure}
    \centering
    \includegraphics[width=\linewidth]{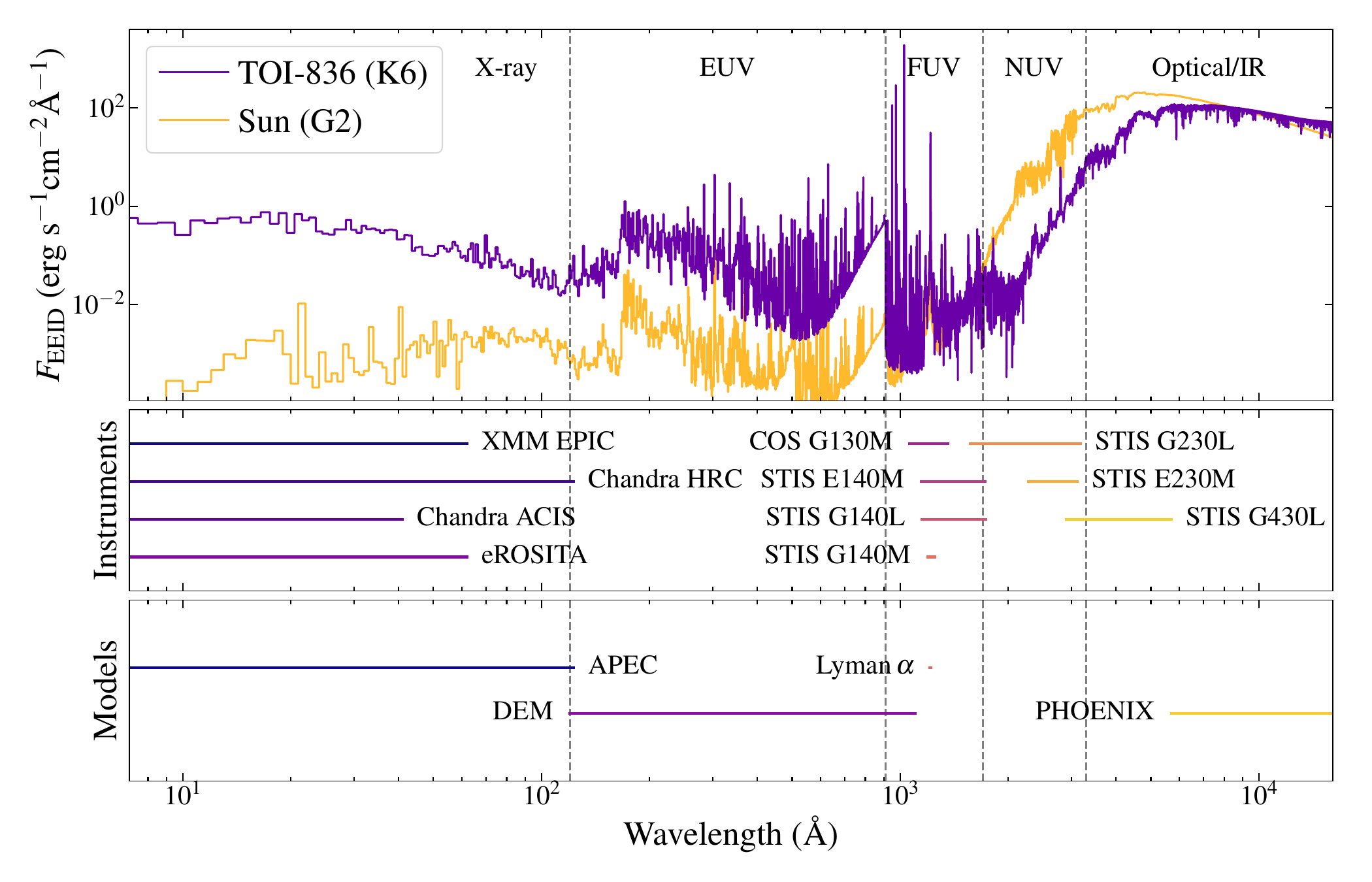}
    \caption{Top panel: SED of TOI-836, scaled to the Earth Equivalent Instellation Distance \citep[EEID][]{mamajek+stapelfeldt23-1} and compared with the Sun. The dashed lines show the wavelength regions discussed in this paper. Middle panel: Approximate wavelength regions covered by the various telescope, instrument and grating setups used in this work. Where multiple cenwaves are available only ones used here are shown. Bottom panel: Approximate wavelength regions covered by model spectra. For both observations and models the exact wavelength regions used in each SED vary slightly depending on (for example) the quality of the data.}
    \label{fig:toi836_sed}
\end{figure}

\section{SED construction}\label{sec:sed}
\begin{deluxetable}{lcccccccc}
\tablecaption{Target list and stellar parameters. Stars in this and all following tables are given in order of increasing \Teff, restarting the order for the four proxy stars. \label{tab:tab_targs}}

\tabletypesize{\small}
\tablecolumns{9}
\tablehead{\colhead{Star} & \colhead{Spectral Type} & \colhead{\Teff\ (K)} & \colhead{Radius (R$_{\odot}$)} & \colhead{Mass (M$_{\odot}$)} & \colhead{$P_{\mathrm{Rot}}$ (days)} & \colhead{Distance (pc)} & \colhead{Av (mag)} & \colhead{Refs}} 
\startdata
GJ\,4102 & M3.5V & $3300^{+80}_{-30}$ & $0.2789\pm0.0014$ & $0.284\pm0.025$ & $79.32$ & 12.48 & \nodata & 1 \\
K2-18 & M2.5V & $3457\pm39$ & $0.411\pm0.038$ & $0.359\pm0.047$ & $38.6^{+0.6}_{-0.4}$ & 38.1 & \nodata & 2 \\
GJ\,367 & M1 & $3522\pm70$ & $0.457\pm0.013$ & $0.454\pm0.011$ & $51.3\pm0.13$ & 9.42 & \nodata & 3, 4 \\
TOI-776 & M1V & $3709\pm70$ & $0.538\pm0.024$ & $0.544\pm0.028$ & $30-40$ & 27.15 & \nodata & 5 \\
GJ\,341 & M0 & $3770\pm40$ & $0.5066^{+0.0169}_{-0.0172}$ & $0.48\pm0.03$ & 7.9--15.0 & 10.45 & \nodata & 6 \\
TOI-134 & M1V & $3800\pm70$ & $0.6\pm0.022$ & $0.62\pm0.03$ & $29.8\pm1.3$ & 25.18 & \nodata & 7, 8 \\
TOI-260 & M0V & $4026\pm14$ & $0.607\pm0.014$ & $0.616\pm0.032$ & $31\pm6$ & 20.21 & \nodata & 10 \\
TOI-178 & mid/late K & $4316\pm70$ & $0.651\pm0.011$ & $0.65^{+0.027}_{-0.029}$ & $\leq36$ & 62.81 & \nodata & 11 \\
NGTS-10 & K5V & $4400\pm100$ & $0.697\pm0.036$ & $0.696\pm0.04$ & $17.29\pm0.008$ & $\approx300^a$ & $0.0366^{+0.074}_{-0.04}$ & 12 \\
TOI-836 & K7V & $4552\pm154$ & $0.665\pm0.01$ & $0.678^{+0.049}_{-0.041}$ & $21.987\pm0.097$ & 27.51 & \nodata & 13 \\
K2-141 & K7V & $4599\pm79$ & $0.681\pm0.018$ & $0.708\pm0.028$ & $13.9\pm0.2$ & 62.0 & $0.14^{+0.14}_{-0.1}$ & 14 \\
HATS-72 & K4 & $4656\pm9$ & $0.7214\pm0.0021$ & $0.7311\pm0.0028$ & $48.725\pm0.015$ & 128.06 & $0.027^{+0.008}_{-0.008}$ & 15 \\
TOI-402 & K1V & $5125\pm50$ & $0.856\pm0.017$ & $0.9\pm0.03$ & $36.5\pm0.2$ & 44.86 & $0.02^{+0.02}_{-0.02}$ & 16 \\
TOI-421 & G9V & $5325^{+79}_{-58}$ & $0.871\pm0.012$ & $0.852^{+0.029}_{-0.021}$ & $43.24^{+0.57}_{-0.55}$ & 74.96 & $0.11^{+0.12}_{-0.08}$ & 17 \\
WASP-63 & G8 & $5550\pm100$ & $1.76^{+0.11}_{-0.08}$ & $1.1^{+0.04}_{-0.06}$ & $\leq31.8$ & 288.94 & $0.11^{+0.04}_{-0.04}$ & 18, 19 \\
HD\,80606 & G5V & $5561\pm24$ & $1.04\pm0.02$ & $1.15\pm0.27$ & $20.0\pm10.0$ & 66.03 & \nodata & 18, 20 \\
Kepler-51 & mid-G & $5670\pm60$ & $0.881\pm0.011$ & $0.985\pm0.012$ & $8.222\pm0.007$ & 801.73 & $0.39^{+0}_{-0.0}$ & 21, 22 \\
HIP\,67522 & G0V & $5675\pm75$ & $1.38\pm0.06$ & $1.22\pm0.05$ & $1.418\pm0.016$ & 124.73 & $0.12^{+0.06}_{-0.06}$ & 23 \\
WASP-166 & F9V & $6050\pm50$ & $1.22\pm0.06$ & $1.19\pm0.06$ & $12.3\pm1.9$ & 114.55 & $0.093^{+0}_{-0.0}$ & 24 \\
WASP-121 & F6V & $6459\pm140$ & $1.458\pm0.03$ & $1.353^{+0.08}_{-0.079}$ & $\approx1.1$ & 272.01 & $0.028^{+0.037}_{-0.022}$ & 25, 26 \\
\textit{Proxy stars} &&&&&& \\
70 Oph\,B & K4V & $4359\pm100$ & $0.67\pm0.008$ & $0.73\pm0.01$ & \nodata & 5.11 & \nodata & 27, 28 \\
$\epsilon$\,Indi & K5V & $4754\pm35$ & $0.736\pm0.007$ & $0.76\pm0.01$ & $32.9\pm0.07$ & 3.64 & \nodata & 29, 30 \\
$\tau$\,Ceti & G8V & $5333\pm78$ & $0.816\pm0.012$ & $0.783\pm0.012$ & $34.0$ & 3.65 & \nodata & 31--35\\
$\kappa^1$\,Ceti & G5V & $5709\pm11$ & $0.946$ & $0.98$ & $9.3$ & 9.28 & \nodata & 36--38 \\
\enddata
\tablecomments{References: (1)\,\citet{lustig-yaegeretal23-1}; (2)\,\citet{cloutieretal17-1}; (3)\,\citet{goffoetal23-1}; (4)\,\citet{lametal21-1}; (5)\,\citet{luqueetal21-1}; (6)\,\citet{ditomassoetal25-1}; (7)\,\citet{astudillo-defruetal20-1}; (8)\,\citet{mannetal19-1}; (10)\,\citet{hobsonetal24-1}; (11)\,\citet{leleuetal21-1}; (12)\,\citet{mccormacetal20-1}; (13)\,\citet{hawthornetal23-1}; (14)\,\citet{malavoltaetal18-1}; (15)\,\citet{hartmanetal20-1}; (16)\,\citet{gandolfietal19-1}; (17)\,\citet{carleoetal20-1}; (18)\,\citet{stassunetal17-1}; (19)\,\citet{bomomoetal17-1}; (20)\,\citet{llorente-de-andresetal21-1}; (21)\,\citet{libby-robertsetal20-1}; (22)\,\citet{johnsonetal17-1}; (23)\,\citet{rizzutoetal20-1}; (24)\,\citet{hellieretal19-1}; (25)\,\citet{bourrieretal20-1}; (26)\,\citet{delrezetal16-1}; (27)\,\citet{morrisetal19-1}; (28)\,\citet{eggenbergeretal08-1}; (29)\,\citet{laliotisetal23-1}; (30)\,\citet{soubiranetal22-1, soubiranetal24-1}; (31)\,\citet{baliunasetal96-1}; (32)\,\citet{breweretal16-1}; (33)\,\citet{vonbraun+boyajianetal17-1}; (34)\,\citet{teixeiraetal09-1}; (35)\,\citet{tang+gai11-1}; (36)\,\citet{rucinskietal04-1}; (37)\,\citet{mamajek+stapelfeldt23-1}; (38)\,\citet{guinanetal03-1}.} 
\tablenotetext{a}{The distance to NGTS-10 is poorly constrained due to an unreliable Gaia parallax (Appendix \ref{sec:notes}).}
\end{deluxetable}


\subsection{Overview}
Following the methodology of MUSCLES and similar surveys, our SEDs combine observations in wavebands accessible with current facilities with model spectra filling in the gaps. The ideal collection of observations consists of a Chandra or XMM X-ray spectrum and FUV and NUV spectroscopy with HST, including, where necessary, a dedicated observation of the \ion{H}{1}~\lya\,1215\,\AA\ emission line. These observations are then combined with APEC thin-plasma models and Differential Emission Measure (DEM) models filling in the gap between the X-ray and FUV spectra, a reconstruction of the \lya\ line that removes the effects of ISM absorption, and a PHOENIX stellar atmosphere model covering the optical and infrared. An example SED and the wavebands covered by the instruments and models used is shown in Figure \ref{fig:toi836_sed}\footnote{Plots throughout the paper scale the SEDs to the Earth Equivalent Instellation Distance \citep[EEID=1\,au $\sqrt{\frac{L_*}{L_\odot}}$,][]{pecaut+mamajek13-1}, allowing at-a-glance comparison with the Sun as seen from Earth.}.

Table \ref{tab:tab_targs} shows our target list, along with relevant stellar parameters. The parameters were taken from the literature, using the NASA exoplanet archive \citep{christiansenetal25-1}\footnote{\url{https://exoplanetarchive.ipac.caltech.edu/}} to collate the most up-to-date parameters. Because the SEDs are created primarily for planetary modeling, we use the stellar parameters presented in the planet discovery papers wherever possible. 

In some cases we were unable to obtain data in the X-ray and/or FUV due to some or all of the targets being either too faint, having too high an ISM column density along the line of sight, or too bright (and thus violating HST bright object protection limits). For stars without complete observations we have replaced the missing regions with data from carefully selected proxy stars.

The following subsections describe the production of the SEDs, starting with the observational data and followed by the models. Full details of all of the observations and notes on the SEDs and (if applicable) proxies for individual stars be found in Appendix \ref{sec:notes}.

\begin{figure}
    \centering
    \includegraphics[width=0.5\linewidth]{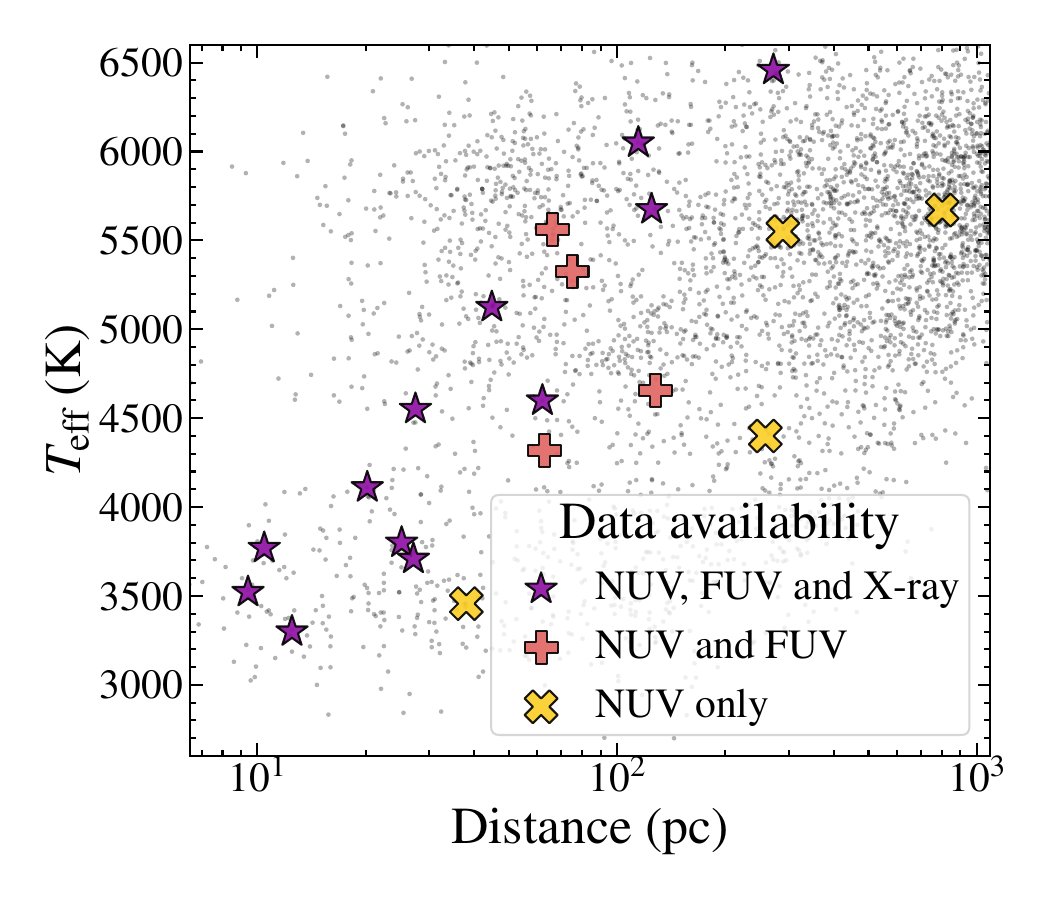}
    \caption{Distance and effective temperatures of the sample, colored according to whether or not the stars have X-ray and/or FUV observation (either new to this program or archival). There were no stars for which we could obtain X-ray detections but no FUV spectrum. The background scatter plot shows the distance and \Teff\ for all exoplanet host stars within the plot limits, retrieved from the NASA exoplanet archive. }
    \label{fig:which_data}
\end{figure}

\begin{deluxetable}{lcccccc}
\tablecolumns{4}
\tablewidth{0pt}
\tablecaption{Parameters for the APEC models used in the SEDs. \label{tab:xray}}
\tablehead{\colhead{Star} &
            \colhead{$F_{\mathrm{x}}$ (10$^{-14}$ erg s$^{-1}$ cm$^{-2}$)} &
            \colhead{kT1 (keV)} & \colhead{kT2 (keV)} & \colhead{Abundance (Solar)} &
            \colhead{$\log$\,N(\ion{H}{1}) (cm$^{-2}$)} & \colhead{Telescope/Instrument}
                  }
\startdata
GJ 4102 & $<0.22$ & $0.3^a$ & \nodata & 0.4 & \nodata & XMM/EPIC \\
GJ367 & $10.0\pm0.8$ & $0.12^{+0.09}_{-0.02}$ & $0.67^{+0.07}_{-0.13}$ & 0.4 & 18.0 & XMM/EPIC \\
TOI-776 & $2.5\pm0.5$ & $0.12\pm0.03$ & $0.83\pm0.13$ & 0.4 & 18.0 & XMM/EPIC \\
GJ341 & $4.1^{+0.5}_{-0.45}$ & $0.14^{+0.04}_{-0.03}$ & $0.83^{+0.27}_{-0.17}$ & 0.4 & 18.5 & XMM/EPIC \\
TOI-134 & $2.4\pm0.3$ & $0.19\pm0.026$ & $0.9^{+0.46}_{-0.5}$ & 0.4 & 18.0 & XMM/EPIC \\
TOI-260 & $1.41^{+0.45}_{-0.34}$ & $0.15^{+0.06}_{-0.04}$ & $0.77^{+0.3}_{-0.2}$ & 0.4 & 19.0 & XMM/EPIC \\
TOI-178 & $0.23\pm0.12$ & $0.24^b$ & \nodata & 0.4 & 19.0 & Chandra/HRC \\
TOI-836 & $7.6^{+1.0}_{-0.9}$ & $0.26\pm0.025$ & $\leq0.7$ & 0.4 & 17.0 & XMM/EPIC \\
K2-141 & $1.37^{+0.28}_{-0.36}$ & $0.28^{+0.03}_{-0.05}$ & \nodata & 0.4 & 19.0 & XMM/EPIC \\
TOI-402 & $1.38\pm0.16$ & $0.193^b$ & \nodata & 0.6 & 19.0 & Chandra/HSC \\
HD 80606 & $<0.07$ & $0.3^a$ & \nodata & 0.4 & \nodata & XMM/EPIC \\
HIP 67522 & $122.0\pm8.0$ & $1.13\pm0.07$ & \nodata & 0.4 & 19.0 & Chandra/ACIS \\
WASP-166 & $0.78\pm0.17$ & $0.43^b$ & \nodata & 0.6 & 18.0 & Chandra/HRC \\
WASP-121 & $1.4\pm0.5$ & $0.39\pm0.0001$ & \nodata & 0.4 & 20.7 & XMM/EPIC \\
$\tau$\,Ceti & $60.0\pm20.0$ & $0.8^{+0.3}_{-0.1}$ & \nodata & 0.4 & 17.0 & XMM/EPIC \\
$\epsilon$\,Indi & $164.0^{+25.0}_{-19.0}$ & $0.22\pm0.03$ & $1.1^{+0.6}_{-0.3}$ & 0.4 & 17.0 & eROSITA \\
\enddata
\tablenotetext{a}{Fixed to fit upper limits.}
\tablenotetext{b}{Assumed values.}

\end{deluxetable}

\subsection{X-ray}\label{sec:xrayobs}
The X-ray region of the SEDS ($\lambda \lesssim 120$ \AA) are based on a combination of new and archival observations and/or estimates from upper limits. XMM and/or Chandra archival data were available for 6 targets, and eROSITA data were available for the proxy star $\epsilon$\, Indi (Appendix \ref{sec:moreseds}). For the remainder, we used the relationship between rotation period and X-ray flux from \citet{wrightetal18-1} to predict the expected count rate and obtain observations with either the EPIC instrument on board XMM-Newton for brighter targets ($\gtrsim 300$ counts expected in the pn detector) or the HRC-I instrument on Chandra for fainter targets ($\gtrsim 30$ counts expected)\footnote{Chandra ACIS is unfortunately no longer sensitive enough for observations of faint soft X-ray sources, meaning that X-ray spectral information cannot be obtained for many targets.}. Six targets were found to be too faint for either observing mode. 

X-ray spectra from XMM/EPIC, eROSITA and archival Chandra/ACIS observations were fit using the {\sc XSPEC} package \citep[v12,][]{arnaud96-1} with Astrophysical Plasma Emission Code \citep[APEC,][]{smithetal01-1, fosteretal12-1} models with 1--2 temperatures combined with an absorption profile from a fixed interstellar hydrogen column density typically $N_{\mathrm{H}} = 10^{18-19}$\,cm$^{-2}$. Column densities below a few times $10^{19}$ cm$^{-2}$ do not affect the derived fluxes significantly: For example, changing the column density from $10^{18}$ cm$^{-2}$ to $10^{19}$ cm$^{-2}$ affects the 0.2 -- 2.4 keV flux by only a few percent. The models were used to fill in the gaps between the highest wavelength of the X-ray spectra (typically $\approx 50$\,\AA) and 120\,\AA, as well as to adjust the XMM spectra for photon losses as described in \citet{wilsonetal21-1}.  

The Chandra HRC observations provide an X-ray count rate over the waveband 1--123\,\AA\ (0.1--10\,keV) but no spectral information. Background-subtracted X-ray count rates were measured using {\sc ciao} version 4.14 \citep{fruscioneetal06-1}. APEC models with parameters taken from similar stars \citep[e.g.][]{ayres+buzasi22-1, brownetal23-1} were then scaled to the count rate to estimate the X-ray flux and spectrum, and used in the SED to fill the entire range 1--120\,\AA. Uncertainties on all X-ray models were estimated by generating 10000 models with parameters randomly drawn from a normal distribution based on the fitted value and 1\,$\sigma$ uncertainties for each parameter, then using the standard deviation of the resulting flux values for each wavelength bin.

\subsection{Ultraviolet and Optical Spectroscopy}\label{sec:uvobs}
The targets were observed with HST using the Space Telescope Imaging Spectrograph (STIS) and, for some targets, the Cosmic Origins Spectrograph (COS) instruments under program ID \pid. FUV spectra were obtained with either STIS G140L (1150--1730\,\AA) or COS G130M with the 1222\,\AA\ cenwave (1064--1360\,\AA). Where practical, we also obtained STIS G140M spectra at the  1222\AA\ cenwave, providing medium-resolution data covering the \lya\ line. In the NUV, we obtained G230L spectra covering 1570--3180\,\AA, as well as an optical G430L spectrum covering 2900--5700\,\AA. Archival data in these or similar modes was used instead of new observations where available. 

All observations were retrieved from MAST after they had been processed using the default {\sc CalCOS} and {\sc CalSTIS} data reduction pipelines. In several cases the automated pipeline failed to identify the spectral trace, in which case the traces were identified by eye in the flt images and re-extracted using {\sc stistools}. Spectra using the same instrument setup were coadded with an uncertainty-weighted average. As with previous surveys we did not remove contributions of flares in the spectra. The only star to exhibit strong flares during the observations was HIP 67522, for which we present a detailed analysis in Froning et al. (submitted). Each spectrum was then visually inspected to determine what range to include in the SED, removing poor quality pixels at the edges of the spectra and determining which spectra should be included in regions of overlap. In almost all cases the red end of a lower-waveband spectrum was used instead of the overlapping blue end of a higher waveband spectrum. 

The ultraviolet spectrum of stars can be significantly affected by interstellar reddening at larger distances. We retrieved literature measurements for stars (Section \ref{sec:phx}) and, where necessary, corrected the spectra using the astropy {\sc dust\_extinction} \citep{gordon24-1} implementation of the \citet{fitzpatrick99-1} model. However, in most cases the reddening was consistent with zero, as expected for targets in the Local Bubble.  

\begin{figure}
    \centering
    \includegraphics[width=\linewidth]{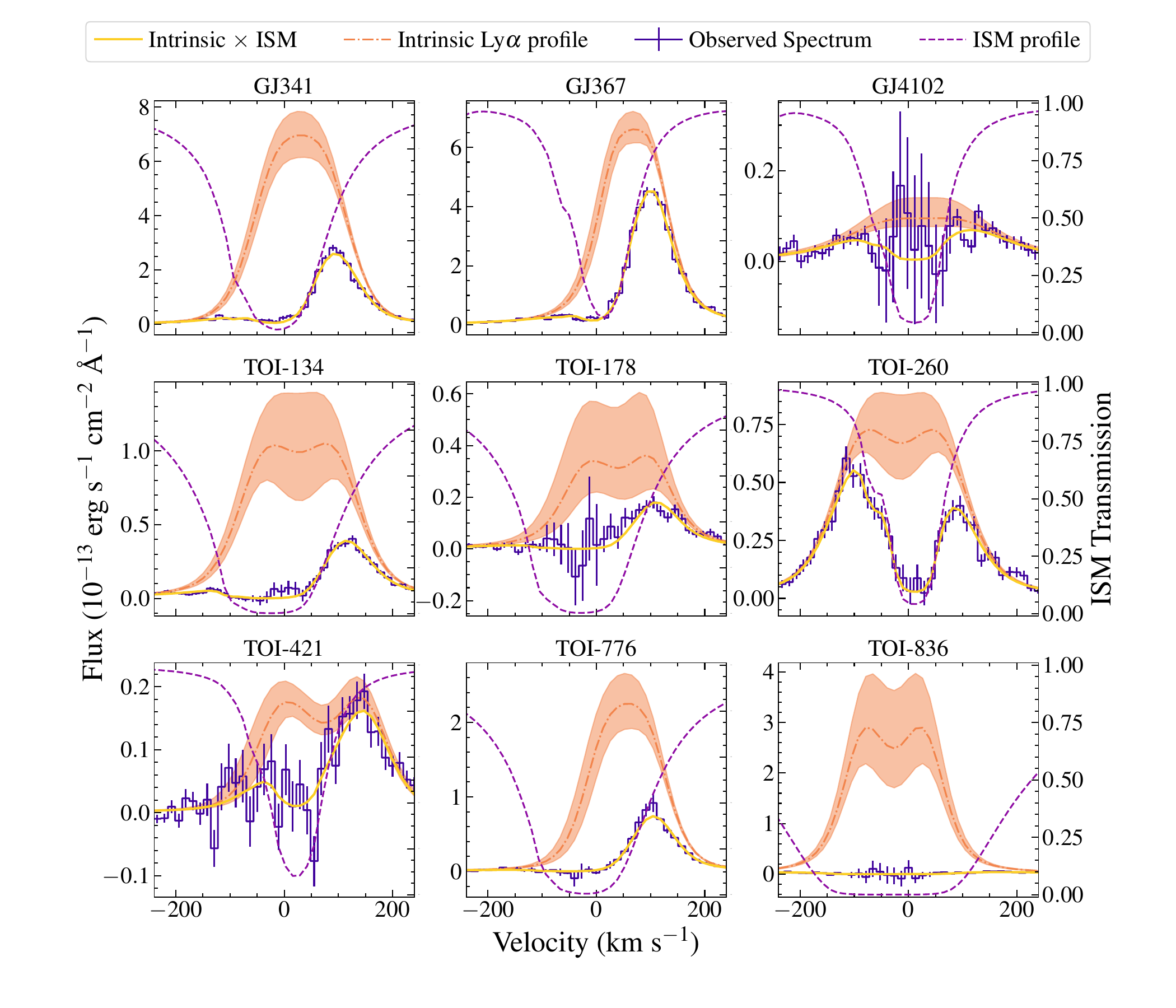}
    \caption{New \lya\ reconstructions for this work, based on STIS G140M data. The orange dash-dotted line is the reconstructed intrinsic emission line, with the shaded area showing the $\pm1\,\sigma$ uncertainties. Convolving the intrinsic line with the ISM profile (purple dashed line) results in the yellow model, reproducing the data (purple). Regions of the data with high uncertainties (e.g. in the line center of GJ\,4102, top right) track where the spectrum was affected by airglow and generally do not contribute to the fit. Stars not shown here used either previously published reconstructions or scaling relationships from other emission lines and/or rotation period.}
    \label{fig:lya}
\end{figure}

\subsection{Lyman alpha reconstructions}\label{sec:lya}
\ion{H}{1} \lya\ (1215\,\AA) is by far the brightest FUV emission line, but absorption from the ISM and contamination from geocoronal airglow makes observing the full profile impossible for most stars. Slit spectroscopy with STIS is used to remove the airglow. We then take two approaches to recover the \lya\ line profile: reconstruction from the broad wings or an estimate based on scaling relations between activity indicators.  In stars with low column density and/or high intrinsic line flux, the broad wings of the emission line are detected and are used to reconstruct the full line profile using {\sc lyapy} \citep{youngblood+newton22-1}. In short, a model of the stellar intrinsic profile attenuated by interstellar \ion{H}{1} and \ion{D}{1} gas is convolved with the instrumental line spread function and forward-modeled to the observed spectrum (see Equations 1-3 from \citealt{youngbloodetal22-1}). In the case of the absence of data or a poorly constrained reconstruction, the intrinsic flux of the line is estimated using scaling relationships with other ultraviolet emission line fluxes or stellar rotation period \citep{woodetal05-1,youngbloodetal16-1, melbourneetal20-1}. A Gaussian profile with full width half max 150 km s$^{-1}$ is then assumed in order to complete the SED. The estimated intrinsic fluxes and relevant fitted parameters are reported in Table~\ref{tab:lya}.

\begin{deluxetable}{cccc}
\tablecolumns{4}
\tablewidth{0pt}
\tablecaption{Parameters of the \lya\ line reconstructions or estimates for the sample. \label{tab:lya}}
\tablehead{\colhead{Star} &
            \colhead{$F_{\mathrm{Ly\alpha}}$ (10$^{-14}$ erg s$^{-1}$ cm$^{-2}$ \AA$^{-1}$)} & 
            \colhead{$\log$\,N(\ion{H}{1}) (cm$^{-2}$)} & 
            \colhead{Self reversal parameter $p$ $^a$}
                  }
\startdata	
GJ 4102 & $1.32^{+0.38}_{-0.16}$ & $17.84^{+0.32}_{-0.24}$ & \nodata \\
K2-18$^{b}$  & 3.2$^{+3.1}_{-2.6}$ & 18.16$^{+0.44}_{-0.34}$ & \nodata  \\ 
GJ 367  & $42.9^{+3.9}_{-3.2}$ & 17.78$\pm$0.09 & 1.4$\pm$0.2 \\
TOI-776  & $15.3^{+2.8}_{-2.3}$ & 18.65$\pm$0.06 & 1.2$^{+0.2}_{-0.1}$  \\
GJ 341  & $54.6^{+6.0}_{-5.6}$ & 18.37$^{+0.05}_{-0.06}$ & 1.1$^{+0.2}_{0.1}$ \\
TOI-134  & 11.0$^{+2.8}_{-2.2}$ & 18.71$^{+0.07}_{-0.08}$ & 1.6$^{+0.5}_{-0.4}$ \\
TOI-260  & 8.56$^{+0.13}_{-0.10}$ & 17.80$^{+0.13}_{-0.14}$ & 1.6$\pm$0.4 \\
TOI-178  & 3.53$^{+0.20}_{-0.12}$ & 18.60$^{+0.22}_{-0.36}$ & 1.9$^{+0.4}_{-0.3}$ \\
NGTS-10$^c$  & 0.49$\pm$0.11 & \nodata & \nodata \\
TOI-836$^{d}$  & 26.0$^{+7.7}_{-5.5}$ & 19.08$\pm$0.17 & 2.0$\pm$0.3 \\
K2-141$^c$  & 9.13$\pm$1.66 & \nodata & \nodata \\
HATS-72$^c$  & 1.22$\pm$0.28 & \nodata & \nodata \\
TOI-402$^c$  & 16.3$\pm$3.6 & \nodata & \nodata \\
TOI-421  & 1.91$^{+0.36}_{-0.23}$ & 17.72$^{+0.34}_{-0.40}$  & 2.4$\pm$0.1 \\ 
WASP-63$^e$  & 0.17 & \nodata & \nodata \\ 
HD 80606$^c$  & 6.89$\pm$5.65 & \nodata & \nodata \\
Kepler-51$^e$  & 0.12 & \nodata & \nodata \\
HIP 67522$^c$  & 58.6$\pm$11.9 & \nodata & \nodata \\
WASP-166$^c$  & 5.25$\pm$1.79 & \nodata & \nodata \\
WASP-121$^e$  & 1.16$\pm$0.03 & \nodata & \nodata \\
\enddata
\tablenotetext{a}{$p=0$ corresponds to a full emission line with no depression near the line core, $p=1$ corresponds to enough self-reversal that the line core is flat-topped, and $p>1$ corresponds to a visible dip in intensity near the line core.}
\tablenotetext{b}{Reconstructed flux and H I column density from \cite{dossantosetal20-1}.}
\tablenotetext{c}{\lya\ flux estimated based on the stellar rotation period \citep{woodetal05-1, youngbloodetal16-1}.}
\tablenotetext{d}{Gaussian priors were used for the intrinsic \lya\ parameters to constrain the fit to be roughly consistent with the expected flux of a late K dwarf. This forced the H I column density value to be larger than otherwise expected for a star at 28 pc \citep{woodetal05-1}. }
\tablenotetext{e}{\lya\ flux is scaled from a proxy star (see Table~\ref{tab:proxies}). WASP-63's proxy is the Sun \citep{woodsetal09-1}, Kepler-51's proxy is $\kappa^1$ Cet \citep{woodetal05-1}, and WASP-121's proxy is Procyon \citep{cruzaguirreetal23-1}. }
\end{deluxetable}

\subsection{Differential Emission Measure models}\label{sec:dem}
With the EUV unobservable due to the combination of a lack of suitable observatories and strong interstellar hydrogen absorption, we used differential emission measure (DEM) models to infer the unobserved regions of the high-energy spectrum between the red end of the X-ray data and APEC model (120\,\AA) and the blue end of the HST data (usually $\approx1150$\,\AA). The inputs for the models are the X-ray spectra and/or APEC models combined with FUV emission line fluxes. The emission line fluxes were measured from the HST data using {\sc spec2flux}\footnote{\url{https://github.com/bellalongo/spec2flux}}, a Python routine written for this project. First, the radial velocity of the spectrum is calculated by fitting Voigt profiles to strong emission lines (typically some or all of \ion{Si}{3}\,1206\,\AA\ and the \ion{C}{2}\,1335\AA, \ion{Si}{4}\,1400\,\AA\ and \ion{C}{4}\,1550\,\AA\ doublets). The radial velocity is then used to estimate the positions of weaker lines which may otherwise have been missed or misidentified at their rest wavelengths. The line positions of a list of common ultraviolet lines are then visually inspected, with a Voigt fit attempted. For each line, we then determine whether to integrate over the model fit or over the spectrum following the methodology of \citet{franceetal18-1}, or if the line is a non-detection. We find that for a given line, fluxes calculated from a model fit and from integrating over the spectrum are consistent, but that fitting first lowers the uncertainty. In either case a continuum subtraction is performed by fitting a polynomial to a short region on either side of the line. If the line is not detected we report it as a $3 \sigma$ upper limit. Line lists for targets with FUV spectra are given in the Appendix (Table \ref{tab:linelist}).      

These flux measurements and upper limits were combined with either the X-ray spectra or APEC models as inputs to the DEM implementation developed in \citet{duvvurietal21-1} and \citet{duvvurietal23-1}.  The DEM uses the assumption of coronal equilibrium to simplify emission from plasma as an integral of the product of two functions: the contribution function, which describes the energy emitted by a volume of collisionally excited plasma as a function of temperature specific to each emission line and/or recombination continuum and which is calculable using atomic data (\texttt{CHIANTI v10}, \citealp{Dere:1997_CHIANTI_I_V1, DelZanna:2021_CHIANTI_XVI_V10}) and the coronal equilibrium assumption, and the DEM, which quantifies the amount of collisionally excited plasma along the line of sight as a function of temperature. The \citet{duvvurietal21-1} implementation uses the contribution functions of observed lines and spectra, a parameterization of the base-10 logarithm of the DEM as a fifth-order Chebyshev polynomial, and a parameterization of the model's systematic uncertainty as an average fraction of the predicted flux to fit for the range of DEM functions consistent with the observations, then combines this envelope of DEM possibilities with the contribution functions of emission lines and hydrogen/helium continua between 120 -- 1150\,\AA\ to predict an ensemble of $10^4$ EUV spectra given both the observational and model systematic uncertainties. The spectrum in the SED takes the median of the flux at each wavelength from this ensemble and the difference between the median and 16th and 84 percentile values to calculate asymmetric uncertainties. The logarithmic mean of the uncertainties is used for for the final symmetric error array in the HLSP.

\subsection{PHOENIX optical and infrared models}\label{sec:phx}
Wavelengths red-wards of the STIS G430L spectra ($\gtrsim5690$\,\AA) use a PHOENIX BT-Settl (CIFIST)\citep{allard16-1} model. The model grid was retrieved from the SVO Theoretical spectra web server\footnote{\url{http://svo2.cab.inta-csic.es/theory/newov2/}} and  converted into vacuum wavelengths, then interpolated to match the stellar parameters in Table \ref{tab:tab_targs}. Uncertainties were estimated by generating models with \Teff$\pm \sigma_{\mathrm{\Teff}}$ and taking the average difference in flux between them and the original model for each bin (see \citealt{wilsonetal25-1} for a longer discussion).

We encourage users of these SEDs to replace the PHOENIX models with their own data and/or models generated using stellar parameters of their choice, such that any analysis remains internally consistent \citep[see for e.g. ][]{thaoetal24-1}. The G430L spectra were taken with the wide (52X0.2) slit for minimal slit losses and generally show excellent agreement with other absolutely flux-calibrated data, so can be used as a baseline to scale alternative optical/IR spectra onto the rest of the SED. We note that the wavelength coverage of the JWST instruments extends over much of the wavelength space covered by the PHOENIX models, such that a full, space-based optical/IR SED for several stars may be built up over time as a happy side effect of exoplanet observations.

\begin{deluxetable}{lccccccc}
\tablecaption{Summary of stars where proxies were used. Stellar parameters are either copied from Table \ref{tab:tab_targs} or taken from the given references. In the third column, ``$<$'' indicates that the proxy star was used for the entire SED bluewards of the given wavelength.  \label{tab:proxies}}

\tabletypesize{\small}
\tablecolumns{8}
\tablehead{\colhead{Star} & \colhead{Proxy} & \colhead{Proxy Wavelengths (\AA)} & \multicolumn{2}{c}{\Teff (K)} & \multicolumn{2}{c}{$P_{\mathrm{rot}}$ (days)} & \colhead{Reference}\\
&&& \colhead{Star}& \colhead{Proxy} & \colhead{Star} & \colhead{Proxy} &}
 
\startdata
K2-18 & GJ\,832 &$<1699$ & $3457\pm39$ & $3539^{+74}_{-79}$ & $38.6^{+0.6}_{-0.4}$ & $40\pm5.1$ & \citet{loydetal16-1}, this work\\
GJ\,367 & GJ\,674 & 1367--1715 &  $3522\pm70$ & $3404^{+57}_{-59}$ & $51.3\pm0.13$ & 32.9 & \citet{wilsonetal25-1}\\
GJ\,341 & GJ\,676A & 1367--1624 & $3770\pm40$ & $4014^{+90}_{-94}$ & $7.9-15$ & 41.2& \citet{wilsonetal25-1} \\
TOI-134 & GJ\,649 & 1364--1699 & $3800\pm70$ & $3621^{+41}_{-40}$ & $29.8\pm1.3$ & 23.8 & \citet{wilsonetal25-1} \\
TOI-260 & GJ\,410 & 1365--1649 & $4111\pm171$ & $3786^{+83}_{-89}$ & $31.0\pm6$ & 14.0 & \citet{pinedaetal21-2} \\
NGTS-10 & 70 Oph B & $<1214$, 1217.5--1650 & $4400\pm100$ & $4359\pm100$ & $17.29\pm0.008$ & \nodata & This work \\
HATS-72 & $\epsilon$\,Indi & $<1214$, 1217.5--2100& $4656\pm9$ & $4754\pm35$ &$48.725\pm0.015$ & $32.9\pm0.07$ & This work \\
TOI-421 & $\tau$\,Ceti & $<1159$ & $5325^{+79}_{-58}$ & $5333\pm78$  & $43.24^{+0.57}_{-0.55}$ & 34.0 & This work \\
WASP\,63 & Sun & $<1980$ & $5550\pm100$ & $5772\pm0.8$ &$\leq31.8$ & $\approx25$ & \citet{woodsetal09-1}   \\ 
Kepler\,51 & $\kappa^1$\,Ceti & $<1699$ & $5670\pm60$ & $5709\pm11$ & $8.222\pm0.007$ & 9.3 & This work\\
WASP-121 & Procyon & \lya\ & $6459\pm140$ & $6474\pm94$ & $\approx1.1$ & $23\pm2$ & \citet{cruzaguirreetal23-1} \\
\enddata
\end{deluxetable}

\subsection{Proxies}\label{sec:prox}
For ten targets we were unable to obtain the full observational dataset required to build an SED, so proxy stars were required to build out the complete SED (Table \ref{tab:proxies}). For each star we searched for stars with similar spectral type and rotation period that either have published, publicly available SEDs or sufficient archival data to construct one ourselves. Details of the four additional targets that we ended up constructing are given in Appendix \ref{sec:moreseds}. The proxies were either scaled by distance or to shared spectra and/or photometry. In all cases we retained as much as possible of the original target spectrum (instead of, for example, replacing areas where the target was observed but the proxy had better S/N). The choice of proxy and scaling methods by star are given in Appendix \ref{sec:notes}. If a proxy DEM is used then the ultraviolet fluxes used to calculate the DEM will only have come from the proxy star, even if FUV data for the target is available (e.g in the case of TOI-421). 

As mentioned above, significant differences have been found in the X-ray and ultraviolet spectra for stars with similar characteristics, and even for the same star observed at different stages in its activity cycle \citep[e.g.][]{kamgaretal24-1}. Using proxies may therefore introduce a considerable systematic error into models using these SEDs, and we recommend caution when reporting any results that rely on a proxy spectrum. Comparing with additional stars of a similar spectral type that have data in the wavebands of interest is a sensible precaution.

\section{Discussion}\label{sec:disc}
Our targets were drawn from the list of transiting planet systems observed during JWST Cycle 1, providing a sample that is not statistically representative or unbiased. Our ability to draw broad conclusions about stellar astrophysics is therefore limited, so instead we will discuss the implications of the stellar SEDs for the specific planets being observed and the outlook for host star SED availability in JWST Cycle 2 and beyond.  

\subsection{Mega-MEATS and the Sun}
\label{sec:sun}
\begin{figure}
    \centering
    \includegraphics[width=\linewidth]{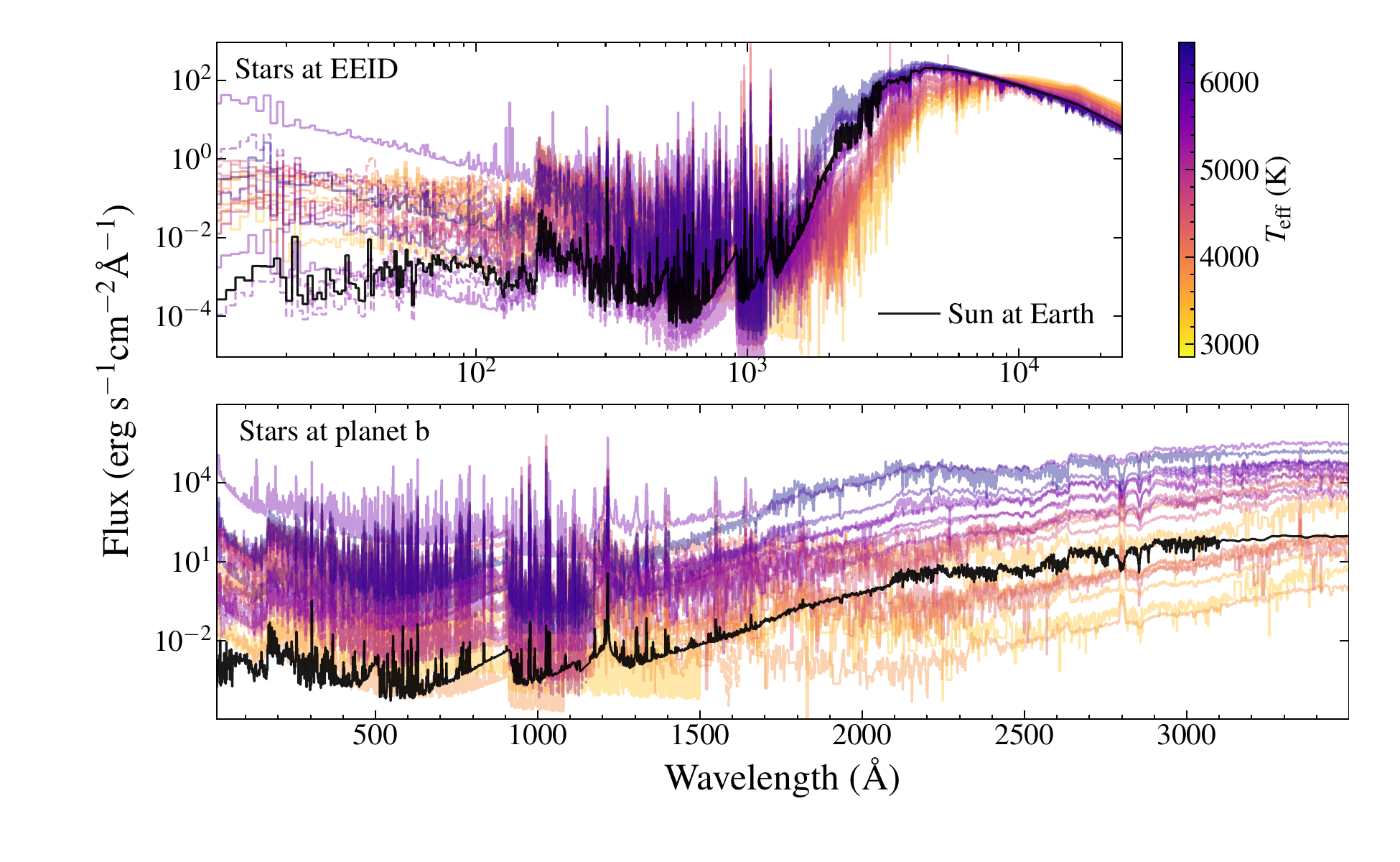}
    \caption{SEDs scaled to the EEID (top panel) and to the semi-major axis of each planet b (bottom panel). In the bottom panel the wavelength range is changed to expand the X-ray and ultraviolet range. The Sun at 1\,au is overplotted in black for reference in both panels. Dashed lines show regions where proxies were used.}
    \label{fig:stars_eeid}
\end{figure}
\begin{figure}
    \centering
    \includegraphics[width=\linewidth]{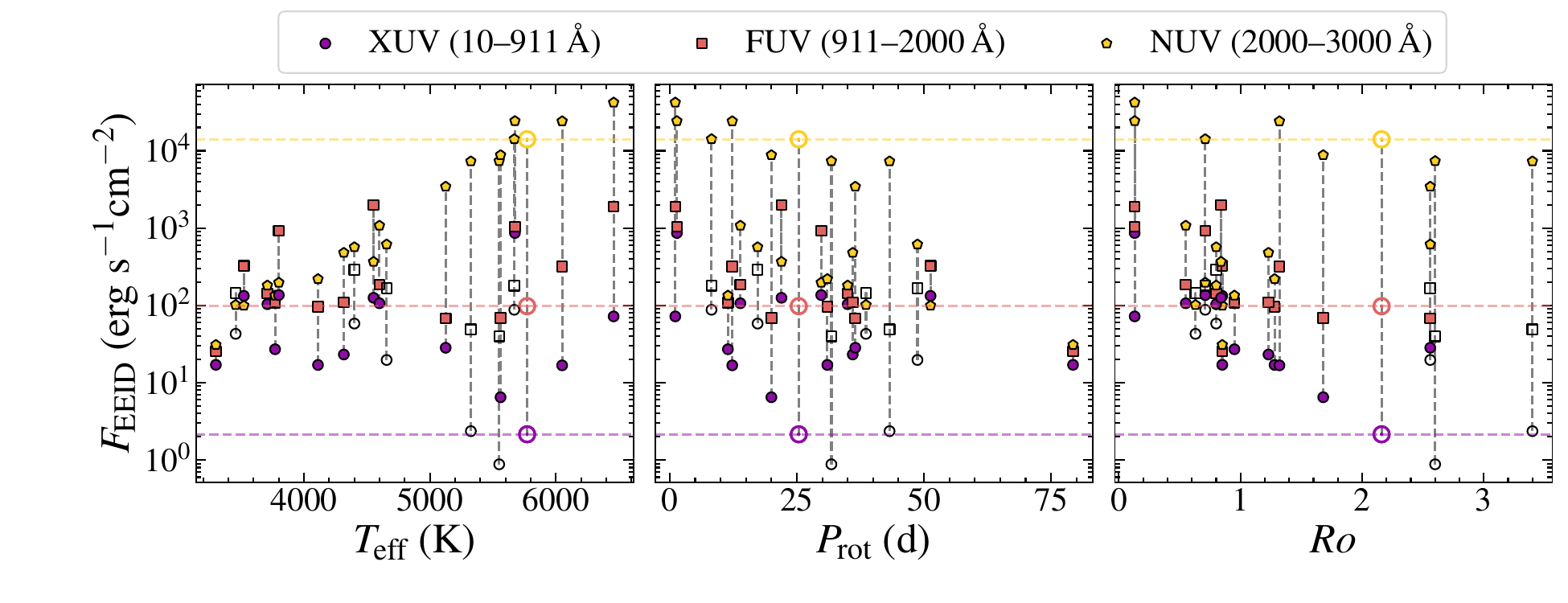}
    \caption{Integrated fluxes in three wavebands as a function of stellar \Teff, rotation period and Rossby number. Points for the same star are joined by vertical dashed lines. Solar values were found by integrating the \citet{woodsetal09-1} spectrum and are shown with the $\bigodot$ symbol. The dashed horizontal lines extend the Solar values across the x-axis for easy comparison with the other stars. Empty markers show fluxes derived mainly from proxies.}
    \label{fig:stars_tpr}
\end{figure}

\begin{figure}
    \centering
    \includegraphics[width=0.5\linewidth]{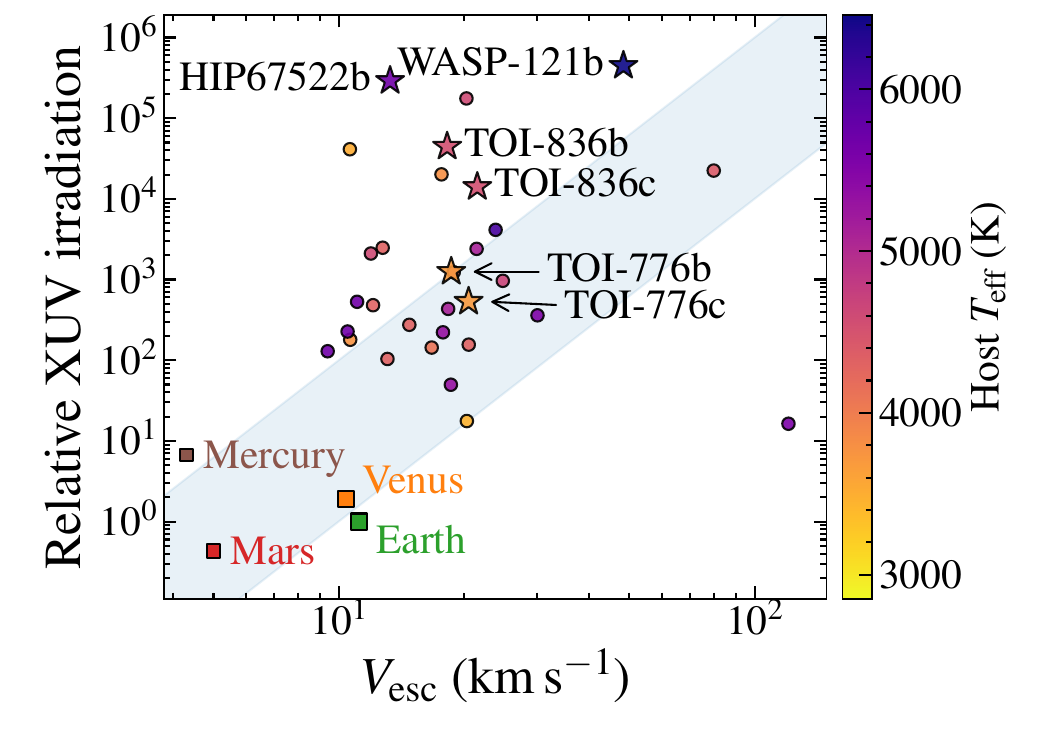}
    \caption{Instantaneous XUV irradiation relative to the Earth compared with escape velocity for the planets orbiting our stars with known masses and radii. The shaded area shows the order-of-magnitude range around the by-eye cosmic shoreline from \citet{zahnle+catling17-1}. Planets with detected atmospheric escape are marked with stars and labeled. }
    \label{fig:shoreline}
\end{figure}

Figure \ref{fig:stars_eeid} shows the entire sample scaled to the EEID (top panel) and to the distance of each system's planet b, which is the planet observed in the majority of JWST programs (bottom panel), and colored according to effective temperature. The SEDs here and in following plots are compared to the quiet Solar SED assembled by \citet{woodsetal09-1}\footnote{\url{https://lasp.colorado.edu/lisird/data/whi_ref_spectra}}. Apparent from the top panel is that the XUV\footnote{X-ray and EUV, defined here as 10--911\,\AA. We cut off the integration at 10\,\AA\ as not all SEDs have data below that.} flux experienced in the habitable zones of these planets is decoupled from spectral type, with orders-of-magnitude XUV flux difference between stars of similar \Teff. In both plots the Earth has among the lowest XUV instellation, something particularity notable in the bottom panel where the XUV instellation of the Earth is low compared with the exoplanets despite middle-of-the-pack NUV instellation. XUV-driven atmospheric modification, expansion and/or escape likely plays a larger role at JWST target planets than it currently does in the Solar system, even at those planets with a similar total instellation to the Earth. 

This finding does assume that we have estimated the currently unobservable EUV fluxes correctly. The EUV flux typically contributes 5-10 times the flux of the observable X-ray flux to the total XUV flux (see Table \ref{tab:euvfluxes} for a full breakdown), and is the stronger driver of atmospheric escape \citep[e.g.][]{youngbloodetal25-1}. Whilst the DEM models represent the state-of-the-art EUV flux estimates and can reproduce the few available stellar EUV spectra \citep{duvvurietal25-1}, other EUV estimates are available based on scaling relationships from, for example, X-ray flux \citep{sanz-forcadaetal11-1}, FUV emission line fluxes \citep{franceetal18-1}  and \lya\ flux \citep{linskyetal14-1}. We find that our EUV fluxes are comparable with these estimates, although they typically have $\sim$ order-of-magnitude uncertainties so the comparison is limited (The typical S/N of the integrated DEM fluxes is $\approx10$). Further testing of the EUV fluxes is limited by the lack of empirical data, which must be provided by a future space mission.

Figure \ref{fig:stars_tpr}shows integrated XUV, FUV and NUV fluxes as functions of effective temperature, rotation period and Rossby number (calculated using Equation 5 from \citealt{wrightetal18-1}).Stars where the EUV and FUV spectra are proxies are shown with empty markers. Note that, given the close match between target and proxy \Teff\ and $P_{\mathrm{rot}}$ (Table \ref{tab:proxies}) the positions of the points would not noticeably change if the proxy star were plotted instead. Habitable zone XUV and FUV fluxes show no correlation with temperature and rotation period, with some adjacent stars differing by an order of magnitude. NUV fluxes do increase with temperature as blackbody emission from the photosphere shifts into the ultraviolet, an effect that can also be seen in the uptick of FUV fluxes for the hottest stars. The fluxes do generally trend down with increasing Rossby number, especially in the EUV, but still with plenty of scatter. We do not provide a fit to the EUV v Rossby number trend because, as noted above, the sample is not representative so the trend may not be generalisable to other stars. The (lack of) trends remain even if the proxy points are disregarded. Overall, the weak relationships between high-energy fluxes and more easily measurable stellar parameters reinforces the requirement for direct observations of exoplanet host stars and/or carefully chosen proxies.

The \citep{woodsetal09-1} Solar spectrum used here represents the Sun at Solar minimum. To check that the XUV excess remains over the entire Solar cycle, we retrieved spectra for Solar maximum and minimum from the FISM2 database\citep{chamberlinetal20-1}, which provide semi-empirical models of the Solar X-ray and ultraviolet spectrum for every day since 1947\footnote{\url{https://lasp.colorado.edu/lisird/data/fism_daily_hr}. Specifically, we used the Solar spectra for 2020~January~01 and 2024~August~01, which we selected as representative of typical fluxes during Solar minimum and maximum respectively based on the light curves provided by the database.}. We found that the integrated XUV flux of the Sun changed by a factor of $\approx$2.5 over the last Solar cycle, not enough to significantly change its position relative to the other stars on Figures \ref{fig:stars_eeid} and \ref{fig:stars_tpr}. However, this does raise the point that the stars here have been observed at random points in their own activity cycles (if they have them), and similar changes in flux over time could affect their exoplanets. 
If some or all of these systems remain popular JWST targets, then a subset of these observations should be repeated in a few years time to see if the SEDs have changed enough to impact the interpretation of the exoplanet observations.

\subsection{An instantaneous Cosmic Shoreline}

The generally high XUV fluxes experienced by the JWST targets implies that many of them are susceptible to ongoing atmospheric escape. Figure \ref{fig:shoreline} shows a version of the cosmic shoreline from \citet{zahnle+catling17-1} which compares the XUV fluxes with escape velocity\footnote{\citet{ditomassoetal25-1} find a mass upper limit of 4\,\mearth\ for the Venus-sized GJ\,341\,b from radial velocity measurements, which (as they acknowledge) is clearly unphysical. We therefore set the mass to 0.8\,\mearth\ to calculate the escape velocity in Figure \ref{fig:shoreline}}, inspired by the empirical observation in the Solar system that objects with low escape velocities and high XUV fluxes are less likely to have retained atmospheres. Most formulations of the shoreline \citep[e.g.][]{passetal25-1, bertathompsonetal25-1} use an estimated time-integrated XUV flux, which suggest which planets may or may not have retained an atmosphere over their lifetimes. Here we instead show the ``measured'' (observed X-ray plus DEM-modelled EUV) instantaneous XUV fluxes, which indicate which planets may be undergoing atmosphere escape \textit{now}. Highlighted are planets with published detections of atmospheric escape: WASP-121b \citep{singetal19-1}; TOI-836 b and c \citep{zhangetal25-1}; and TOI-776 b and c \citep{loydetal25-1}, as well as HIP\,67522\,b where there is strong circumstantial evidence for ongoing escape \citep{thaoetal24-1}. Four of the planets with escaping atmospheres are, as expected, well above the shoreline, although the picture may be biased by unpublished non-detections. Three more planets without detected escape fall into a similar parameter space. K2-141\,b, receiving $\sim 10^6$ times the XUV instellation of the Earth, is a rocky ultra-short period planet with a likely molten surface and hints of an atmosphere formed by rock vapor \citep{ziebaetal22-1}. Atmospheric formation and escape processes in such an extreme environment are likely quite complex and heavily influenced by the stellar ultraviolet input \citep{nguyenetal22-1}, although \citet{ito+ikoma21-1} find that mineral atmospheres can survive even under high XUV fluxes for rocky planets more massive than Earth. GJ\,376\,b (XUV irradiation $4\times10^4$ Earth) is, as put by \citet{zhangetal24-1}, a ``Dark, Hot, Airless Sub-Earth''\footnote{Presenting a considerable challenge to makers of exoplanet travel posters.}, and the non-detection of an atmosphere is entirely consistent with the cosmic shoreline framework. TOI-134\,b (= L168-9\,b) is a $\approx 4$\mearth\ super-Earth \citep{astudillo-defruetal20-1, hobsonetal24-1} with XUV irradiation $2\times10^4$ that of Earth: any atmosphere is likely long gone but its presence will be tested with JWST observations. The detections of atmospheric escape at the sub-Neptunes TOI-776 b and c \citep{loydetal25-1} are less consistent with the cosmic shoreline framework, and are somewhat in tension with JWST observations returning non-detections of atmospheres at both planets \citep{aldersonetal25-1, teskeetal25-1}. Further planned observations with HST will better characterize the escape signals.

\begin{figure}
    \centering
    \includegraphics[width=\linewidth]{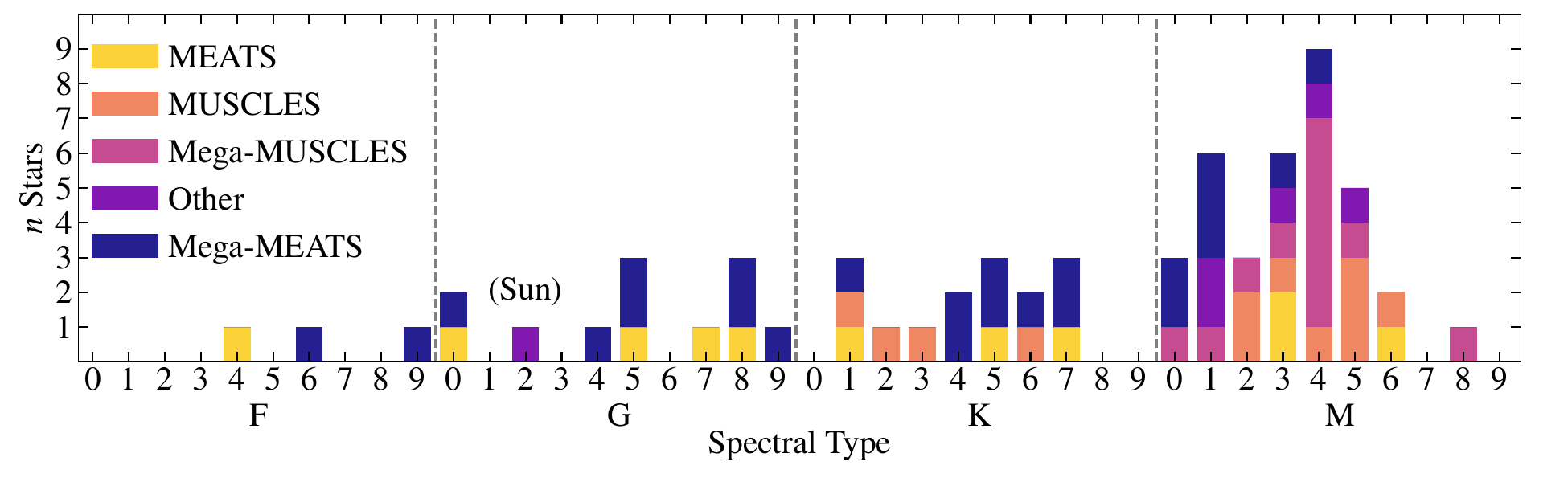}
    \caption{Number of stars with full X-ray to ultraviolet SEDs available as MAST HLSP as a function of spectral type. Fractional spectral types have been rounded up. References: MUSCLES \citep{loydetal16-1}; Mega-MUSCLES \citep{wilsonetal25-1}; MEATS \citep{behretal23-1}; Mega-MEATS (This work); Other \citep{diamond-loweetal21-1, diamond-loweetal22-1, diamond-loweetal24-1, feinsteinetal22-1}.}
    \label{fig:sptypes}
\end{figure}

\begin{figure}
    \centering
    \includegraphics[width=0.5\linewidth]{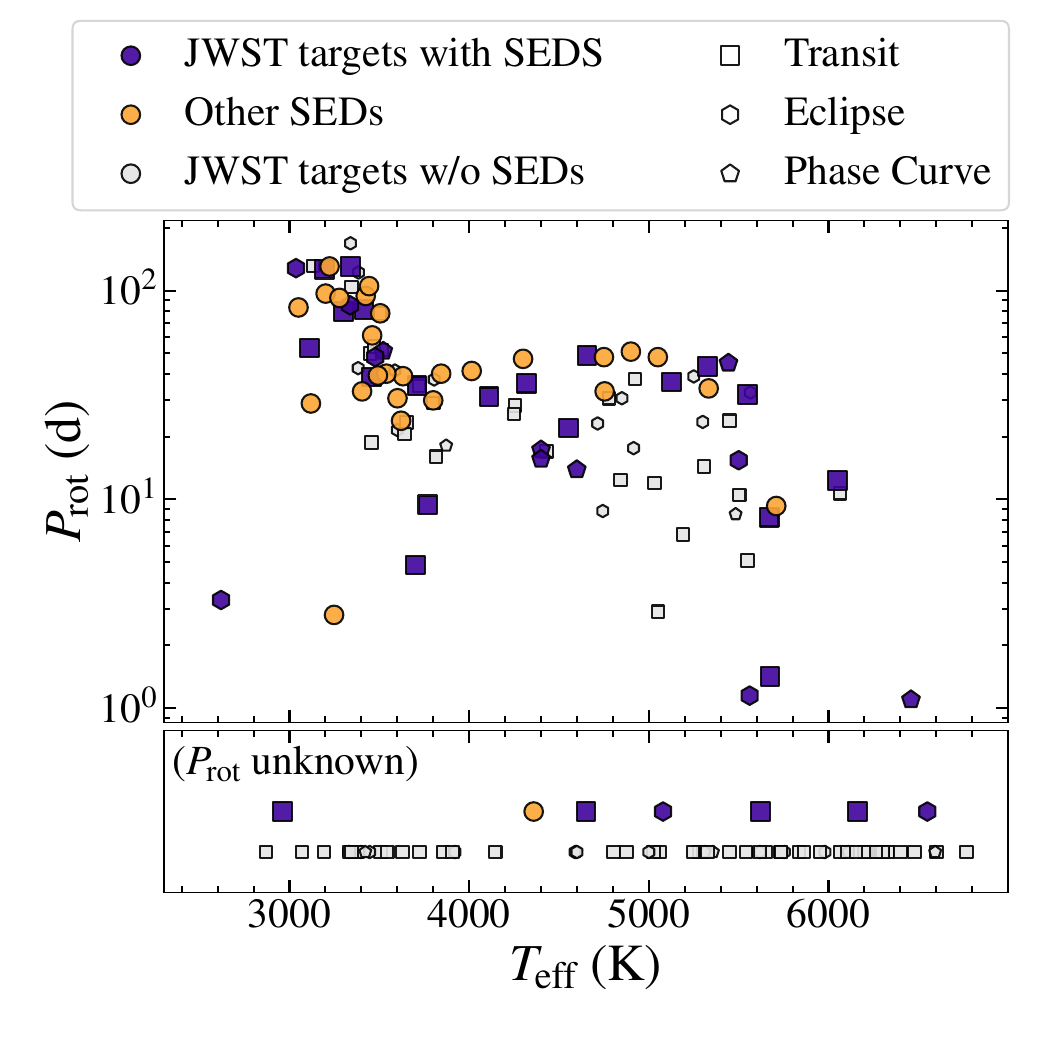}
    \caption{Effective temperatures and rotation periods for all stars targeted for JWST exoplanet observations (except direct imaging) in JWST Cycles 1--4, as well as additional stars with SEDs on MAST. Stellar parameters for the JWST stars are taken from the Exoplanet Archive and from either the references in Figure \ref{fig:sptypes} or \citet{franceetal13-1} for the SEDs. Four stars with \Teff $>8000$\,K are beyond the plot limits.}  
    \label{fig:c1-4}
\end{figure}

\subsection{State of the art for JWST Cycle 2 and beyond}
Mega-MEATS joins an increasing number of programs producing full SEDs for transiting planet host stars. In Figure \ref{fig:sptypes} we summarize the spectral types for which a full SED is available as a MAST HLSP (as well as the \citealt{woodsetal09-1} Solar spectrum). The sample is dominated by early to mid M\,dwarfs, with few F and G stars. M\,dwarf planet hosts will dominate the transiting planet science return for JWST for sub-Neptune and rocky planets, but efforts to fill in the gaps in spectral type will be valuable both to better understand the high-energy behavior of stars across the main sequence and prepare for the Habitable Worlds Observatory (HWO), which will search for Earth analogs around FGK stars \citep{mamajek+stapelfeldt23-1}.  

Is the sample of stellar SEDs sufficient for current and future JWST observations? Figure \ref{fig:c1-4} shows the host star effective temperatures and rotation periods (if known) for all JWST transmission spectroscopy, eclipse and/or phase curve targets in Cycles 1--4, along with the available SEDs. The target list was retrieved from TrExoLiSTS\footnote{\url{https://www.stsci.edu/~nnikolov/TrExoLiSTS/JWST/trexolists.html}} and matched with stellar parameters from the NASA Exoplanet Archive. Most regions of the parameter space have a good selection of nearby SEDs, such that multiple potential proxies are available if direct observations are impractical, although we again advise caution when presenting results that rely on a proxy star \citep[e.g][]{tealetal22-1}. Many of these stars do have some ultraviolet and/or X-ray observational coverage, which could be used to construct partial SEDs with proxies filling in the gaps. There is a notable gap in the coverage of available SEDs around \Teff $\approx5000$\,K, $P_{\mathrm{rot}} \lesssim 10$\,d, where there are several JWST targets without SEDs with similar parameters: constructing SEDs to cover that region should be prioritized in order to maximize the scientific return from JWST.       

Four A-type stars with \Teff $>7000$\,K are beyond the plot limits of Figure \ref{fig:c1-4}, and have no available SEDs or proxies. Three (KELT-20, WASP-178 and KELT-9) have temperatures well in excess of the $\approx8200$\,K point where stars stop forming chromospheres/coronae \citep{guntheretal22-1} and photosphere model atmospheres should suffice to predict high-energy output, although see \citet{fossatietal18-1} for a more detailed discussion. The fourth is WASP-189 which, with \Teff $=8000\pm80$ \citep{lendletal20-1}, may still exhibit coronal emission and should be observed with ultraviolet and/or X-ray telescopes in the future (there are no archival HST COS/STIS, Chandra or XMM observations, although an NUV spectrum was obtained by the CUTE cubesat \citealt{sreejithetal23-1}).     

\section{Conclusions}\label{sec:concs}
We have added SEDs of 20 JWST transiting planet host stars and four nearby stars to the now considerable archive of publicly available stellar SEDs. Combined with previous programs, we now have SEDs or suitable proxies for the majority of exoplanet host stars observed in JWST Cycle 1 and beyond, and are building an increasingly comprehensive understanding of the high energy spectra of F, G, K and M type main sequence stars. We find that the planets in JWST Cycle 1 are receiving XUV fluxes up to six orders of magnitude greater than the Earth, even for planets with similar overall instellation. Under the cosmic shoreline formulation, many of the rocky planets in the sample have likely already lost their atmospheres, with ongoing and potentially detectable escape present at others.       

Given the impact of extreme ultraviolet fluxes on the atmospheres of many of the most observable JWST target planets, the almost complete lack of EUV data for cool stars and resulting reliance on model spectra is a glaring knowledge gap. The MANTIS cubesat \citep{indahl+wilson22-1} will launch the first orbital astrophysics EUV telescope in decades, providing simultaneous EUV, FUV and NUV stellar spectroscopy in concert with JWST observations, but will be limited by the typical short lifetime of cubesats to only a handful of stars. Launching an Explorer- or Probe-class EUV observatory is vital to ensure the best possible science return from JWST exoplanet observations.



\section{Acknowledgments}
We thank the anonymous referee for a helpful review, as well as Tom Ayres and Brian Wood for sharing archival data and useful advice. This work was based on observations made with the NASA/ESA Hubble Space Telescope, obtained from the Data Archive at the Space Telescope Science Institute, which is operated by the Association of Universities for Research in Astronomy, Inc., under NASA contract NAS 5-26555. These observations are associated with program \pid. Support for program \pid\ was provided by NASA through a grant from the Space Telescope Science Institute. All of the HST data presented in this paper were obtained from the Mikulski Archive for Space Telescopes (MAST).
This research has used data obtained with the Chandra X-ray Observatory and from the Chandra Data Archive. The CIAO software package was used to analyze these data. Chandra analysis was supported by SAO grant GO2-23002X. This paper employs a list of Chandra datasets, obtained by the Chandra X-ray Observatory, contained in the Chandra Data Collection (CDC) 513~\dataset[10.25574/cdc.513]{https://doi.org/10.25574/cdc.513}.

This research has made use of data from XMM-Newton, an ESA science mission with instruments and contributions directly funded by ESA member states and NASA. 
This research has made use of data and/or software provided by the High Energy Astrophysics Science Archive Research Center (HEASARC), which is a service of the Astrophysics Science Division at NASA/GSFC. 
This work has made use of data from the European Space Agency (ESA) mission Gaia, processed by the Gaia Data Processing and Analysis Consortium (DPAC). Funding for the DPAC has been provided by national institutions, in particular the institutions participating in the Gaia Multilateral Agreement. 
This work is based on data from eROSITA, the soft X-ray instrument aboard SRG, a joint Russian-German science mission supported by the Russian Space Agency (Roskosmos), in the interests of the Russian Academy of Sciences represented by its Space Research Institute (IKI), and the Deutsches Zentrum für Luft- und Raumfahrt (DLR). The SRG spacecraft was built by Lavochkin Association (NPOL) and its subcontractors, and is operated by NPOL with support from the Max Planck Institute for Extraterrestrial Physics (MPE). The development and construction of the eROSITA X-ray instrument was led by MPE, with contributions from the Dr. Karl Remeis Observatory Bamberg \& ECAP (FAU Erlangen-Nuernberg), the University of Hamburg Observatory, the Leibniz Institute for Astrophysics Potsdam (AIP), and the Institute for Astronomy and Astrophysics of the University of Tübingen, with the support of DLR and the Max Planck Society. The Argelander Institute for Astronomy of the University of Bonn and the Ludwig Maximilians Universität Munich also participated in the science preparation for eROSITA. 
This research has made use of the NASA Exoplanet Archive, which is operated by the California Institute of Technology, under contract with the National Aeronautics and Space Administration under the Exoplanet Exploration Program. This work made use of CHIANTI, a collaborative project involving George Mason University, the University of Michigan (USA), University of Cambridge (UK) and NASA Goddard Space Flight Center (USA).
Y.M acknowledges funding from the European Research Council (ERC) under the European Union’s Horizon 2020 research and innovation programme (grant agreement no. 101088557, N-GINE). 


%

\vspace{5mm}
\facilities{HST (STIS and COS), XMM-Newton, Chandra, eRosita, IUE}

\defcitealias{astropy18-1}{Astropy Collaboration, 2018}
\software{astropy \citepalias{astropy18-1}, XSPEC \citep{arnaud96-1}, stistools\footnote{\url{https://stistools.readthedocs.io/en/latest/}}, scipy \citep{virtanenetal20-1}, numpy \citep{harrisetal20-1}, matplotlib \citep{hunter07-1}, CHIANTI \citep{Dere:1997_CHIANTI_I_V1, DelZanna:2021_CHIANTI_XVI_V10}}



\appendix

\section{Notes on individual stars}\label{sec:notes}

\subsection{GJ 4102}
GJ-4102 was only marginally detected in the XMM observations, so we fitted a nominal 0.3\,keV APEC model to the total count rate and used that to calculate the upper limit on the flux. The DEM was underconstrained at lower temperatures due to the non-detection of \ion{Si}{2} emission in the FUV spectrum, leading to a potential underestimation of flux around the 911\,\AA\ recombination continuum. We compared the DEM flux with that predicted by the \ion{N}{5} scaling relationships from \citet{franceetal18-1} and found them to agree with in a factor $\approx3$; with no additional input to guide us we therefore include the DEM as is, but suggest caution when drawing conclusions that rely on the SED in the range $\approx 800-1000$\,\AA.

\subsection{K2-18}
HST suffered a guide star failure in the first of two visits to K2-18. Comparison of the G230L spectra from the first and second visit showed that after the guide star failure the star was improperly centered on the slit. Around 50\,percent of the flux was lost and an incorrect wavelength solution found. For the G230L spectra, we shifted the wavelength and flux to the correct values by comparison with the (we assume) correct spectrum from the second visit, before coadding all four final spectra. For the G430L spectra, the lines were insufficiently well-resolved to measure the Doppler shift so the wavelength calibration was unchanged. K2-18 has 16 archival Swift U-band observations obtained in 2020~April--July. We measured fluxes in each image with {\sc uvotsource} using a 10" aperture, finding a statistically constant U\,band flux. The G430L spectrum was integrated over the Swift/UVOT U-band filter, and the spectrum flux scaled by the mean of the ratios of the integrated flux to the U-band measurements. 

\citet{dossantosetal20-1} produced a \lya\ reconstruction based on STIS G140M data from HST PID 14221, which we reproduced and included in the SED. Otherwise, for wavelengths $< 1699$\,\AA\ we used GJ\,832 from the MUSCLES survey \citep{loydetal16-1} as a proxy. MUSCLES used the EUV scaling relationships from \citet{linskyetal14-1} instead of DEMs, so we constructed a DEM for GJ\,832 following the methodology described in Section \ref{sec:dem}. The GJ\,832 spectrum was scaled to the same distance as K2-18b, with the new DEM spliced in place of the MUSCLES EUV spectrum\footnote{The GJ\,832 DEM will be added to the MUSCLES HLSP along with this release, but a full update of the MUSCLES SEDs with DEMS has yet to be completed.}. K2-18 was serendipitously observed with XMM in 2024 (Obs. ID 0943530501) and was marginally detected in the EPIC-MOS detectors. We estimate $\log L_{\mathrm{x}} = 26.6\pm 0.4$ in the range 0.2--2.4 keV, in good agreement with the $\log L_{\mathrm{x}} = 26.07\pm 0.6$ found for GJ\,832 by \citet{brownetal23-1}.

\subsection{GJ 367, GJ 341, TOI-134, TOI-260}
These stars, all M\,dwarfs, were found to violate STIS bright object protection polices the case of a flare, so they were observed with the COS G130M grating at the 1222\,\AA\ cenwave to move bright emission lines (\lya, \ion{Si}{4} and \ion{C}{4}) off the detector. We applied post-pipeline adjustments to the COS spectra as described in \citet{wilsonetal25-1}. The COS setup left a gap between $\approx$1370--1700\,\AA, which we filled with STIS G140L taken from stars with similar spectral types and rotation periods from the HST archive. We scaled the proxies to the same distance as the target stars then compared the emission line strengths in the region of overlap to confirm that they were reasonable choices. Emission lines from the proxy spectra were not used as inputs for the DEM.

\subsection{NGTS-10}
The large distance to NGTS-10 made FUV and X-ray observations impractical, so 70\,Oph\,B was chosen as a proxy star with similar spectral type and activity. Stellar parameters for NGTS-10 were obtained from \citet{mccormacetal20-1}, who found that the Gaia DR2 parallax measurement had astrometric noise excess $>2$\,mas and is therefore unreliable. The DR3 measurement has a noise excess of $\approx$1.8, nominally below the reliability limit but close. We found that neither the distance inferred from the DR2 parallax ($325\pm29$\,pc) nor DR3 ($258\pm11$\,pc) reliably scaled the PHOENIX model to the G430L spectrum. We therefore scaled the model directly to the G430L spectrum above 4500\,\AA\ using a least-squares fit. 

Due to the incomplete 70\,Oph\,B spectrum (Appendix \ref{sec:moreseds}) there are no spectral regions of overlap to scale the two SEDs to each other, and as discussed the distance is unreliable. The wavebands of the 70\,Oph\,B G160M and NGTS-10 G230L spectra do overlap between 1568\,\AA\ and 1718\,\AA, but no flux is detected in the G230L spectrum in that range. We therefore used the ratios of the Gaia $G$ magnitudes, as other available photometry for NGTS-10 is blended with a nearby companion \citep{mccormacetal20-1}.

\begin{figure}
    \centering
    \includegraphics[width=\linewidth]{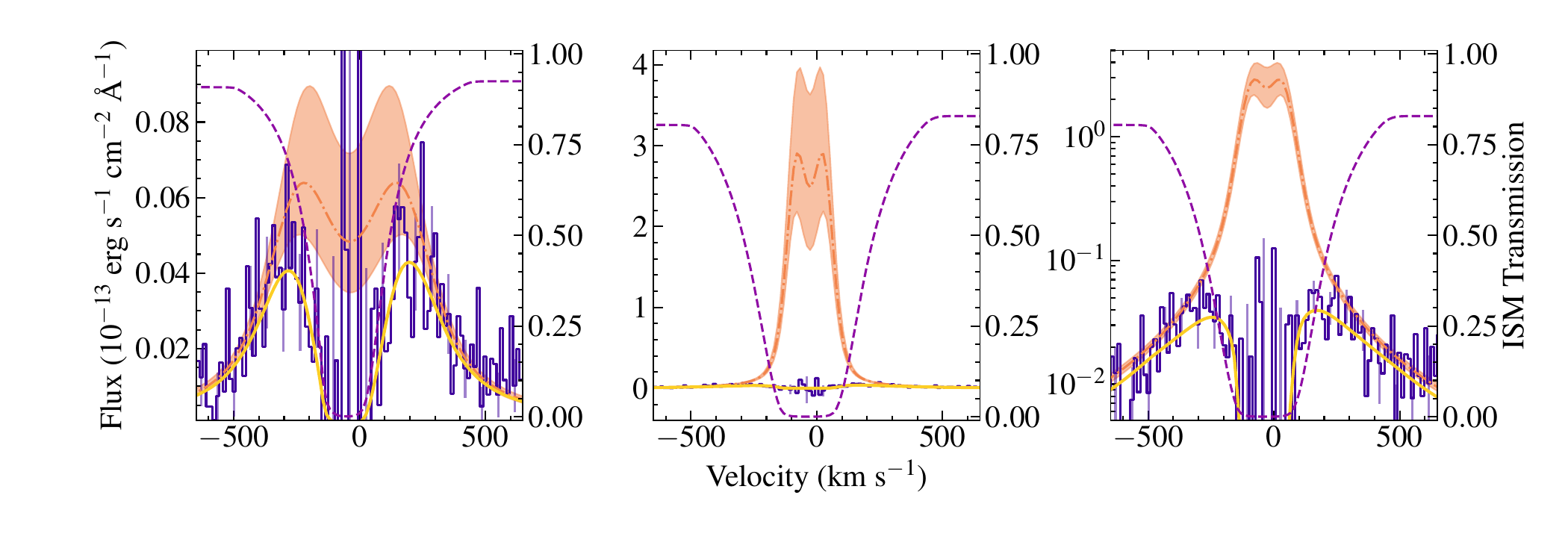}
    \caption{\lya\ profiles for TOI-836 showing the G140M spectrum (purple), reconstructed intrinsic profile (orange), model fit (yellow) and ISM transmission profile (light purple dashed line), as for Figure \ref{fig:lya}. The left panel shows the reconstruction with the default, loose priors on the \lya\ parameters. The middle panel shows the adopted reconstruction with the intrinsic flux constrained to be consistent with similar stars. The right hand panel is the same as the middle but with a log flux scale.}
    \label{fig:toi836lya}
\end{figure}

\subsection{TOI-836}
The \citet{youngbloodetal17-1} rotation period to \lya\ scaling relationship predicts an intrinsic \lya\ flux for TOI-836 of $\approx3.5\times10^{-13}$ erg s$^{-1}$ cm$^{-2}$. An initial {\sc lyapy} reconstruction with the default priors successfully fitted the data (Figure \ref{fig:toi836lya}, left panel), but returned an intrinsic flux of  $\approx2\times10^{-14}$ erg s$^{-1}$ cm$^{-2}$, 17.5 times lower than expected. The reason for the low flux may be that the wings of the line profile, whilst clearly detected, are almost symmetrical, with no information from the line core detected. The symmetric line profile is similar to the zero velocity \lya\ measurement of EG\,UMa by \citet{wilsonetal22-1}, where {\sc lyapy} returned a much weaker line than the (in that case known) true profile. We refit the data, forcing the intrinsic \lya\ flux to be similar to literature values for mid-K dwarfs. We found an intrinsic flux of $2.6^{+0.8}_{-0.6}\times10^{-13}$ erg s$^{-1}$ cm$^{-2}$, still somewhat below the estimate from rotation period but much closer than the unconstrained fit. While the reconstructed line center is now considerably higher flux than the observed spectrum (Figure \ref{fig:toi836lya}  middle panel), the right panel of Figure \ref{fig:toi836lya} shows that the reconstruction is accurately reproducing the wings of the line. However, this solution does require a larger ISM column density (log N(\ion{H}{1}) $=19.08\pm0.17$ cm$^{-2}$) than expected from ISM maps for a star inside 30 pc \citep[e.g $\approx18.4$ cm$^{-2}$]{youngbloodetal25-1}. We do not have a satisfactory resolution to this contradiction, but we retain the constrained reconstruction for the HSLP.

\subsection{K2-141}
The HST observations of K2-141 were affected by a guide star failure. A re-observation was requested to replace a G140L spectrum that was completely lost (dataset oeoo20010), but later analysis showed flux losses in the remaining G140L, G230L and G430L spectra (datasets oeoo20020, oeoo20030, oeoo20040). Photometry from archival Swift UVM2 and U-band images was used to scale the G230L and G430L spectra respectively as described for K2-18 (a single Swift image was available in each band). The second G140L spectrum from the first visit (dataset oeoo20020) was unsalvageable, so only the re-observed dataset oeoo42010 was used in the SED. 

\subsection{HATS-72}
No X-ray data is available for HATS-72 and, despite eight orbits of G140L observations, we did not detect it in the FUV. The star is well-centered in the acquisition images and no guide star issues were reported. We stacked the four flt images from each of two visits (assuming that the trace would fall in roughly the same place on the detector in each visit) but were still unable to identify the spectrum, even at \lya. The most plausible explanation for the non-detection is that the \ion{H}{1} column density is higher than what was estimated when planning the observations. 
 We therefore used $\epsilon$\,Indi as a proxy star for wavelengths $< 2100$\AA, with the exception of \lya\, which was estimated via the scaling relationships to rotation period from \citet{woodetal05-1} and \citet{youngbloodetal16-1}. To scale the proxy we used the $\epsilon$\,Indi E230H spectrum, smoothing it to to the resolution of the HATS-72 G230L spectrum then using a least-squares fit to find the scaling factor. The \ion{Mg}{2} lines in both spectra were masked out during the fit. We estimated an upper limit for the FUV spectrum from the non-detections of $F_{\lambda} \lesssim 4\times10^{-17}$erg s$^{-1}$cm$^{-2}$\AA$^{-1}$ over 1220--1715\,\AA, which the scaled proxy spectrum falls below at all wavelengths. 

\subsection{TOI-421}
Whilst we were able to obtain an FUV spectrum with STIS G140L for TOI-421, the spectrum had low S/N with few recognizable emission lines, and no X-ray data was available. We therefore used $\tau$\,Ceti as a proxy for the X-ray and EUV regions. A simple distance scaling showed good agreement between the NUV spectra of the two stars, so we did not further refine the scaling. 

\subsection{WASP-63}
The large distance and low expected high-energy flux of WASP-63 make X-ray and FUV observations impractical. We therefore filled the wavelength regions below 1980\,\AA\ with the quiet Solar spectrum from \citet{woodsetal09-1}. The Solar spectrum was scaled to match the G230L spectrum between 1980-2600\,\AA\. We added a nominal 10\,percent uncertainty on each flux bin of the proxy spectrum. These uncertainties are mainly included so that codes that require an uncertainty array have one, and should not be taken as accurate estimates of the flux uncertainties of WASP-63.

\subsection{HD 80606}
HD 80606 was also only marginally detected with XMM and we used the same treatment as for GJ 4102.

\subsection{Kepler-51}
Kepler-51 is too distant for FUV and X-ray observations, so we used $\kappa^1$\,Ceti as a proxy. These two stars have practically identical stellar parameters, and scaling the $\kappa^1$\,Ceti to the same distance as Kepler-51 produces an exact match across almost the whole spectrum. There is a troubling exception in the NUV from roughly 2100\,\AA\ (at shorter wavelengths the S/N of the the Kepler-51 spectrum is too low to be usable) to 2550\,\AA, where the scaled $\kappa^1$\,Ceti spectrum underpredicts the Kepler-51 spectrum by nearly a factor of two. Stellar activity should not produce such a strong effect as the photosphere flux dominates at these wavelengths in G stars, the stars have similar metallicites, and we have found no satisfactory instrumental explanation. We chose to scale the proxy spectrum to distance rather than to this region of the NUV spectrum, but users may want to experiment with scaling to the discrepant region. We found that the \ion{Mg}{2}\,2800\,\AA\ lines are completely absorbed by the ISM in the Kepler-51 spectrum, so the region 2790--2810\,\AA\ was also replaced with the scaled $\kappa^1$\,Ceti spectrum.   

\subsection{HIP 67522}
Our attempts to reconstruct the intrinsic \lya\ line of HIP\,67522 returned unphyscial results, discussed extensively in Froning et al. (submitted). Currently the SED uses the \citet{woodetal05-1} scaling relationship version of the \lya\ line discussed in that paper, but we will update the HLSP if future work improves the reconstruction.

\subsection{WASP-121}
WASP-121 has extensive E230M, G430L and G750L observations obtained in the PanCET program \citep{evansetal18-1} which we included in our SED. The orders of the E230M spectra were combined by coadding each overlap region. For the optical spectra we used all of the 100\,s exposure time G430L (12 datasets) and G750L (five datasets) observations. We did not exclude the in-transit spectra as the overall flux difference is only $\sim 1$\,percent. The G430L spectrum obtain in Program \pid\ was not included as the lower exposure time led to no gain in S/N. The red end of the G750L spectra were strongly affected by fringing, which was removed using the {\sc stistools.defringe} routine. 

The combined new and archival spectra have a gap between 1720--2280\,\AA, so we used the PHOENIX model in that range. The model is in good agreement with the observed spectra at both ends: i.e. continuum flux from the chromosphere does not significantly contribute to the NUV spectrum of WASP-121. The G750L spectra extend the observed portion of the SED to 10230\,\AA, with the PHOENIX model covering longer wavelengths as usual.

\subsection{LTT 1445A and GJ 486}
Although LTT 1445A and GJ 486 were observed as part of program \pid, more extensive observations were obtained in HST GO 16722. The program PIs decided to combine efforts, and the analysis and SEDs of these stars are presented in \citet{diamond-loweetal24-1}.

\section{Additional SEDS}\label{sec:moreseds}
\begin{figure}
    \centering
    \includegraphics[width=\linewidth]{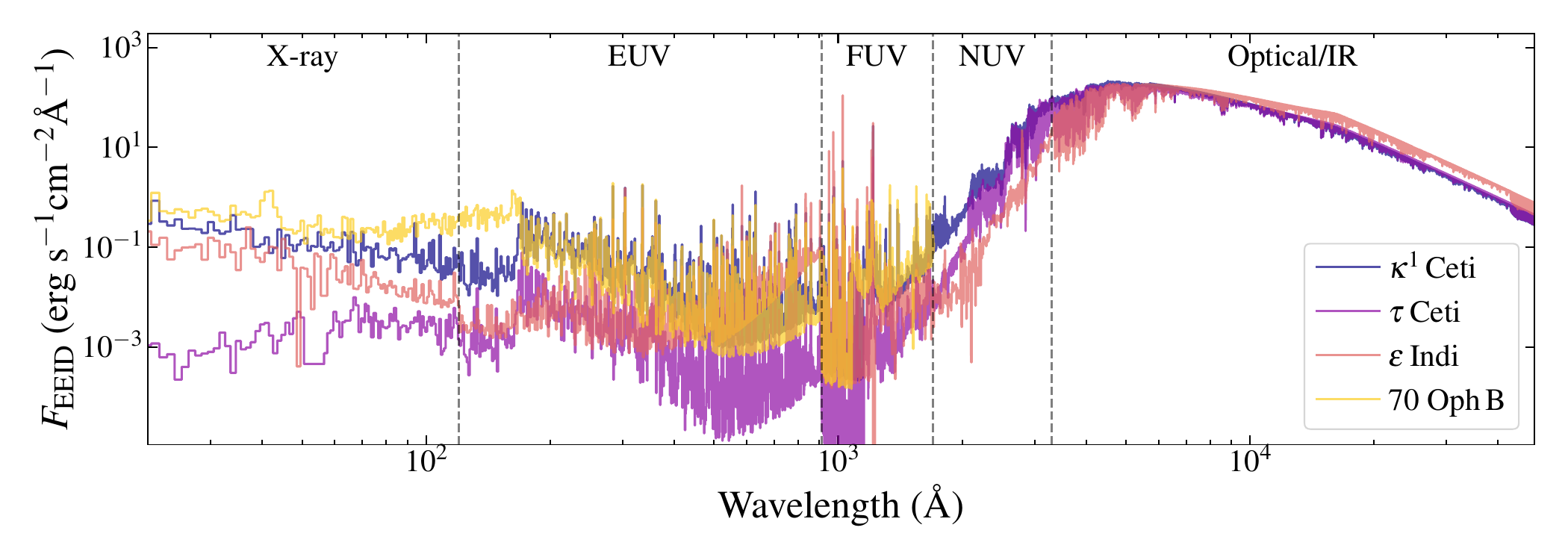}
    \caption{SEDs constructed from archival data used as proxies for stars with insufficient X-ray and/or FUV data.}
    \label{fig:hwostars}
\end{figure}

In several cases we required proxy stars with characteristics that were not represented in the existing set of available SEDs, mainly because our sample focuses on earlier-type stars than previous programs. We therefore have used archival data to produce SEDs of four additional stars. Whilst not JWST targets in Cycle 1, these stars are bright, nearby objects that are high-priority targets for the Habitable Worlds Observatory (HWO), appearing on the ExEP list of likely stars to search for Earth-like planets \citep{mamajek+stapelfeldt23-1}. The SEDs were constructed via the same general approach as the rest of the sample, combining ultraviolet and X-ray spectroscopy with DEM, \lya\ and PHOENIX models. Stellar parameters for the proxy stars were assembled from the literature and are given in Table \ref{tab:tab_targs}. Details specific to each star are given below.
\subsection{$\kappa^1$ Ceti and $\tau$ Ceti}
Archival STIS E140M and E230H spectra are available for both stars, along with E230M and G430L data for $\kappa^1$\,Ceti. For $\kappa^1$\,Ceti we used the the combined echelle spectrum from the STARCAT HSLP\footnote{\url{https://casa.colorado.edu/~ayres/StarCAT/}} \citep{ayres10-1} with the G430L spectrum appended at $\lambda > 2700$\,\AA. For $\tau$\,Ceti we coadded the orders of the echelle observations where they overlapped and filled in the gap between the E140M and E230H spectra (1710--2576\,\AA) with the PHOENIX model. The model was in good agreement with the spectra at both ends of the gap. For $\kappa^1$\,Ceti we used XMM RGS spectra instead of the usual EPIC+APEC model, combining the six available  observations into a single spectrum. \lya\ reconstructions for both stars were reproduced from \citet{woodetal05-1}. 

\subsection{$\epsilon$ Indi}
The $\epsilon$\,Indi SED is by far the most complex assembled here, and provides a proof-of-concept for combining disparate archival data into usable SEDs. We obtained X-ray data from eROSITA DR1 \citep{merlonietal24-1}. All data from the different scans were combined and extracted with a circular source region, and an annulus for the background. The extracted spectrum was then fit with APEC models as for the other stars (Section \ref{sec:xrayobs}). The ultraviolet spectrum was constructed from a Goddard High Resolution Spectrograph (GHRS) spectrum covering 1212--1219\,\AA, International Ultraviolet Explorer (IUE) low resolution spectra covering 1235--2576\,\AA\ and 2835--3350\,\AA\ and a STIS E230H spectrum covering 2576--2835\,\AA. The IUE spectra (29 shortwave and 26 longwave) were combined with a sensitivity-weighted coadd following the algorithm used for the HASP program \citep{debesetal24-1}. We reconstructed the \lya\ line from the GHRS spectrum using {\sc lyapy}.

\subsection{70 Oph B}
Whilst the stellar parameters of 70 Oph\,B make it a good proxy for NGTS-10, the available X-ray and HST data does not allow a complete SED to be reconstructed. Specifically there is no coverage of \lya\ (the available COS spectra being compromised by geocoronal airglow) nor NUV spectra. All STIS spectra in the HST archive labeled as just HD165341 without an A or B suffix are from 70 Oph\,A. We can therefore only assemble a partial SED, but with sufficient coverage to fill in the gaps in the NGTS-10 spectrum. For the X-ray we used the Chandra HRC-S/LETG spectrum from \citet{woodetal18-1} covering 5--175\,\AA. For the FUV we used the available COS G130M and G160M spectra covering 1080-1706\,\AA\ with the region around \lya\ removed. Emission line fluxes from the X-ray and FUV spectra were used to compute a DEM model filling in the range 175--1080\,\AA. A future HST program to complete the SED should be a priority, given the high quality of the existing data and interest in 70 Oph\,B both as a proxy and a HWO target.   

\begin{figure}
    \centering
    \includegraphics[width=\linewidth]{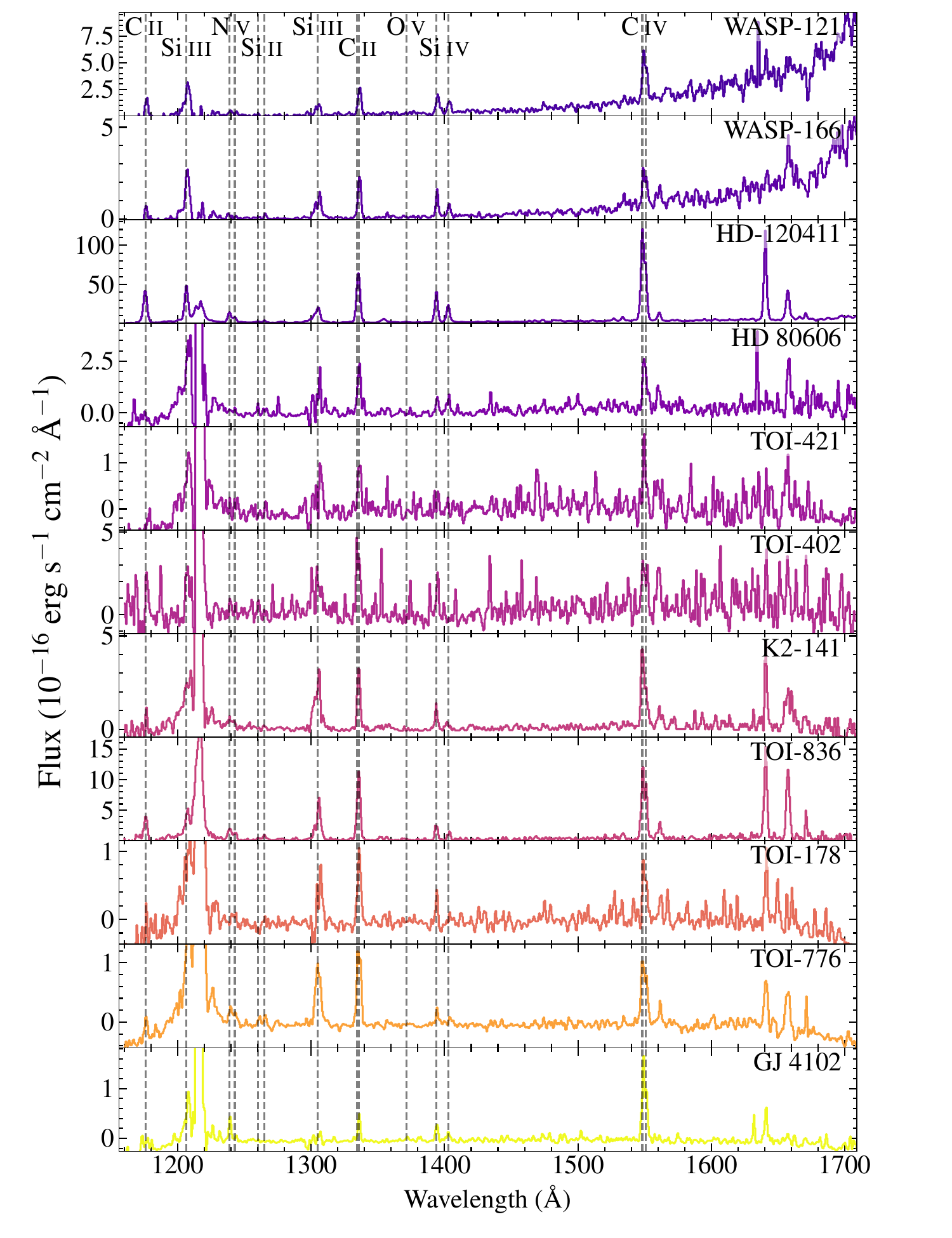}
    \caption{STIS G140L spectra presented here with the major emission lines marked.}
    \label{fig:g140ls}
\end{figure}

\begin{figure}
    \centering
    \includegraphics[width=\linewidth]{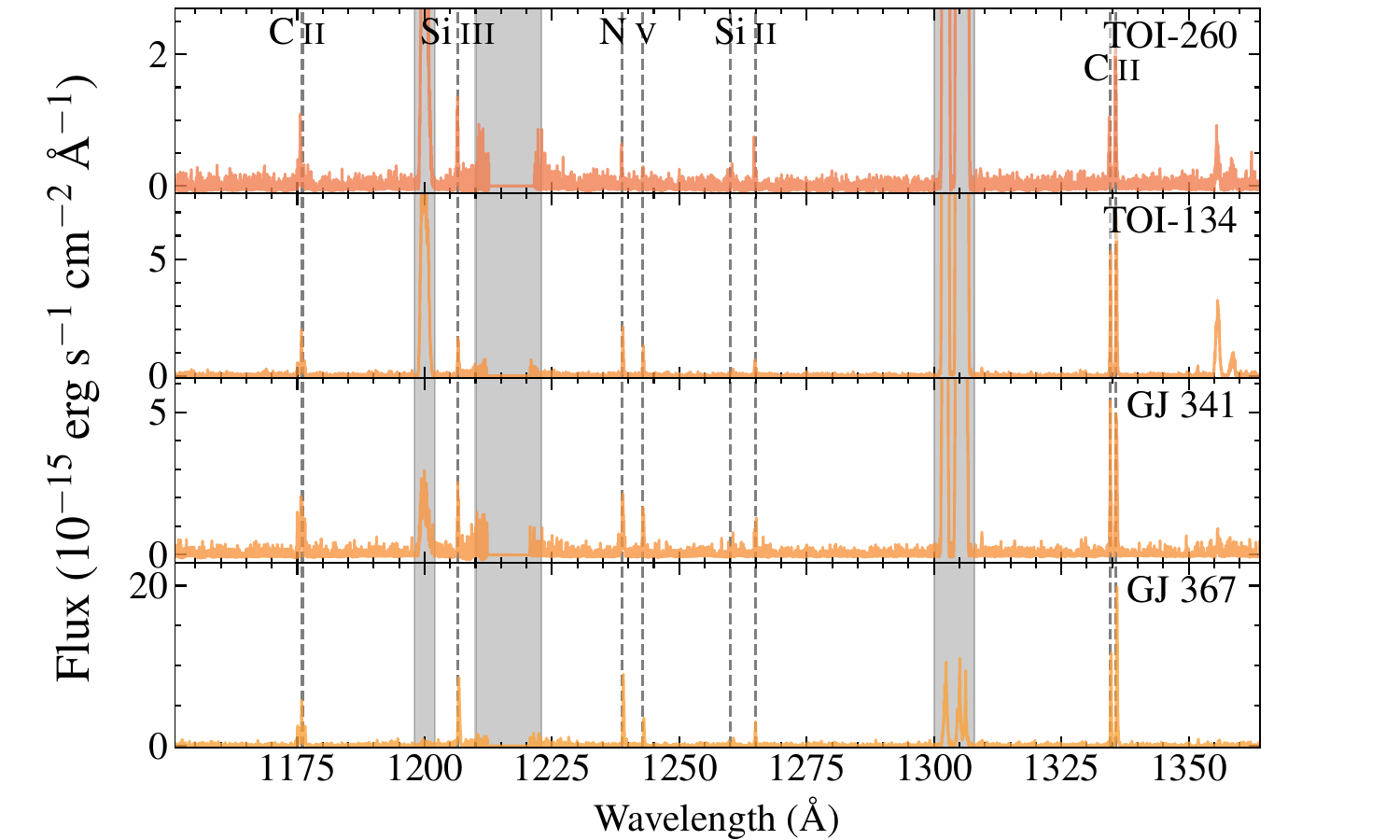}
    \caption{COS G130M spectra. The spectra have been smoothed with a five-point boxcar for clarity. Regions affected by airglow are marked with grey backgrounds. Proxy spectra were used to fill in the rest of the FUV wavelength range for these stars, as detailed in Table \ref{tab:proxies}.}
    \label{fig:g130ms}
\end{figure}

\section{Emission lines in FUV spectra and XUV fluxes}
\label{sec:linelist}
Here we provide plots of the new FUV spectra presented here (Figures \ref{fig:g140ls} and \ref{fig:g130ms}, a list of major emission line fluxes used to  generate the DEM models (Table \ref{tab:linelist}) and integrated X-ray and EUV fluxes (Table \ref{tab:euvfluxes}). The spectra are shown in order of descending \Teff.
\movetabledown=50mm
\begin{rotatetable}


\begin{deluxetable*}{lcccccccccc}
\tablecaption{List of FUV emission line flux measurements used to  create the DEMs. Lines marked ``m'' are multiplets where the given wavelength is the approximate midpoint.\label{tab:linelist}}
\tabletypesize{\small}
\tablecolumns{11}
 \tablehead{\colhead{} & \multicolumn{10}{c}{Line Flux (10$^{-16}$ erg s$^{-1}$ cm$^{-2}$)}\\
 \colhead{} & \colhead{C IIIm} & \colhead{Si III}  & \colhead{N Vm} & \colhead{S II} & \colhead{Si II} & \colhead{Si IIIm} & \colhead{C IIm} & \colhead{O V} & \colhead{Si IVm} & \colhead{C IVm} \\
 \colhead{Star}& \colhead{1175\,\AA} & \colhead{1206.499\,\AA}  & \colhead{1240.0\,\AA}  & \colhead{1260\,\AA} & \colhead{1265\,\AA} & \colhead{1305\,\AA} & \colhead{1335.0\,\AA} & \colhead{1371\,\AA} & \colhead{1400.0\,\AA} & \colhead{1550.0\,\AA}}
\startdata
GJ 4102 & $2.2\pm0.3$ & $3.0\pm0.5$ & $1.0\pm0.2$ & \nodata & $\leq0.3$ & \nodata & $0.9\pm0.2$ & \nodata & $0.8\pm0.4$ & $5.9\pm0.6$ \\
GJ 367 & $26.7\pm1.5$ & \nodata & $27.4\pm1.4$ & \nodata & $5.7\pm0.6$ & $33.1\pm3.1$ & $71.0\pm2.3$ & \nodata & \nodata & \nodata \\
TOI-776 & $2\pm0.35$ & $5.8\pm0.9$ & $2.3\pm0.14$ & \nodata & $0.84\pm0.24$ & $9.7\pm0.2$ & \nodata & $3.1\pm0.34$ & $10.5\pm0.5$ \\ 
GJ 341 & $13.9\pm2.6$ & $6.7\pm1.7$ & $6.0\pm1.5$ & \nodata & $2.9\pm0.7$ & \nodata & \nodata & \nodata & \nodata & \nodata \\
TOI-134 & $6.5\pm0.3$ & $3.9\pm0.4$ & $8.3\pm0.4$ & \nodata & $2.4\pm0.3$ & \nodata & $26.7\pm0.6$ & \nodata & \nodata & \nodata \\
TOI-260 & $3.6\pm0.4$ & $3.5\pm0.4$ & $2.0\pm0.4$ & $\leq1.0$ & $2.0\pm0.4$ & \nodata & $8.9\pm0.5$ & \nodata & \nodata & \nodata \\
TOI-178 & $\leq1.9$ & $\leq3.7$ & $\leq0.9$ & \nodata & $\leq1.1$ & \nodata & $4.2\pm0.4$ & \nodata & $1.5\pm1.1$ & $3.4\pm0.5$ \\
TOI-836 & \nodata & \nodata & $11.5\pm1.2$ & \nodata & $311.0\pm75.2$ & \nodata & \nodata & \nodata & $12.8\pm1.3$ & $49.6\pm3.6$ \\
K2-141 & $\leq4.3$ & $\leq5.8$ & $\leq2.4$ & $\leq1.7$ & $\leq1.5$ & \nodata & $8.8\pm1.0$ & \nodata & $3.6\pm1.5$ & $18.8\pm2.1$ \\
TOI-402 & $\leq11.7$ & $11.5\pm3.1$ & $\leq4.9$ & $\leq4.7$ & $\leq4.5$ & \nodata & $14.7\pm1.7$ & $\leq3.8$ & $14.0\pm4.8$ & $16.0\pm3.0$ \\
HD 80606 & \nodata & $\leq7.5$ & $\leq1.0$ & \nodata & $1.1\pm0.3$ & \nodata & $7.2\pm0.5$ & \nodata & $7.1\pm0.7$ & $11.1\pm0.7$ \\
HIP\,67522 & $213\pm106$ & $216\pm102$ & $85\pm28.5$ & \nodata & \nodata & \nodata & $290\pm115$ & \nodata & $246\pm66$ & $617\pm165$ \\
WASP-166 & $3.7\pm0.4$ & $10.6\pm1.4$ & $1.2\pm0.2$ & $\leq0.6$ & $1.0\pm0.2$ & $6.0\pm0.6$ & $7.1\pm0.3$ & $\leq0.4$ & $5.9\pm0.4$ & $7.8\pm0.8$ \\
WASP-121 & $5.3\pm0.7$ & $14.6\pm1.4$ & $3.0\pm0.4$ & \nodata & $2.8\pm0.3$ & \nodata & $7.4\pm0.5$ & \nodata & $6.6\pm0.5$ & $20.6\pm1.4$ \\
\textit{Proxy Stars} & & & & & & & & & & \\
GJ\,832 & \nodata & \nodata & $31.3\pm0.2$ & $2.2\pm0.1$ & $5.0\pm0.1$ & $0.4\pm0.0$ & $33.4\pm0.2$ & $0.5\pm0.0$ & $22.5\pm0.3$ & $82.4\pm0.5$ \\
70\,Oph\,B & $289.0\pm5.9$ & $231.0\pm6.2$ & $192.0\pm4.7$ & \nodata & $82.5\pm2.8$ & \nodata & $941.0\pm9.6$ & $10.2\pm1.2$ & $260.0\pm5.5$ & $817.0\pm14.0$ \\
$\epsilon$\,Indi & \nodata & \nodata & $200.0\pm100.0$ & \nodata & \nodata & $2370.0\pm225.0$ & $1370.0\pm160.0$ & $\leq529.0$ & $\leq864.0$ & $1560.0\pm281.0$ \\
$\tau$\,Ceti & \nodata & $314\pm9.5$ & $44\pm3$ & $3.2\pm1.0$ & $78.5\pm2.8$ & \nodata & $663\pm7.4$ & $5.0\pm1.4$ & $256\pm4.4$ & $547\pm9.7$ \\
$\kappa^1$ \,Ceti & \nodata & $1810\pm18$ & $95.3\pm3.5$ & \nodata & $160\pm13$ & $224\pm1$ & $1780\pm3$ & $26.3\pm0.4$ & $2140\pm4$ & $843\pm15$ \\
\enddata
\end{deluxetable*}
\end{rotatetable}

\begin{deluxetable}{lcccc}
\tablecaption{Integrated X-ray and EUV fluxes for the sample. \label{tab:euvfluxes}}
\tabletypesize{\small}
\tablecolumns{11}
 \tablehead{\colhead{} & \multicolumn{2}{c}{Flux (10$^{-15}$ erg s$^{-1}$ cm$^{-2}$)}\\
 \colhead{Star}& \colhead{$F_{\mathrm{x-ray}}$ (10-100\,\AA)} & \colhead{$F_{\mathrm{EUV}}$ (100-911\,\AA}  & \colhead{$F_{\mathrm{EUV}}/F_{\mathrm{X}}$}  & \colhead{XUV proxy?} }
\startdata
GJ\,4102 & $0.1\pm0.00$ & $2.0\pm0.04$ & $16.5\pm0.46$ & \\
K2-18 & $0.3\pm0.01$ & $1.2\pm0.08$ & $4.2\pm0.30$ & Y \\
GJ\,367 & $16.0\pm0.97$ & $85.9\pm4.47$ & $5.4\pm0.43$ & \\
TOI-776 & $4.0\pm0.23$ & $12.4\pm0.90$ & $3.1\pm0.29$ & \\
GJ\,341 & $5.6\pm0.14$ & $21.7\pm1.65$ & $3.9\pm0.31$ & \\
TOI-134 & $2.7\pm0.18$ & $31.4\pm2.48$ & $11.5\pm1.19$ & \\
TOI-260 & $1.3\pm0.11$ & $8.4\pm0.77$ & $6.6\pm0.84$ & \\
TOI-178 & $0.3\pm0.12$ & $1.5\pm0.12$ & $4.4\pm1.63$ & \\
NGTS-10 & $0.2\pm0.02$ & $0.2\pm0.01$ & $1.0\pm0.13$ & Y\\
TOI-836 & $8.3\pm0.49$ & $58.7\pm3.04$ & $7.1\pm0.56$ & \\
K2-141 & $1.6\pm0.15$ & $10.6\pm1.10$ & $6.7\pm0.94$ & \\
HATS-72 & $0.1\pm0.01$ & $0.5\pm0.05$ & $4.0\pm0.59$ & Y\\
TOI-402 & $2.0\pm0.70$ & $13.0\pm1.26$ & $6.4\pm2.27$ & \\
TOI-421 & $0.04\pm0.004$ & $0.5\pm0.03$ & $11.3\pm1.21$ & Y\\
WASP-63 & $\approx0.005$ & $\approx0.06$ & $\approx12\pm0.01$ & Y\\
HD\,80606 & $0.1\pm0.01$ & $3.2\pm0.43$ & $48.2\pm9.43$ & \\
Kepler-51 & $0.1\pm0.00$ & $0.1\pm0.01$ & $1.6\pm0.12$ & Y\\
HIP\,67522 & $126.0\pm23.30$ & $116.1\pm7.72$ & $0.9\pm0.18$ & \\
WASP-166 & $0.9\pm0.31$ & $4.5\pm0.49$ & $5.0\pm1.80$ & \\
WASP-121 & $1.5\pm0.20$ & $6.1\pm0.42$ & $4.0\pm0.59$ & \\
&& \textit{Solar max} & $\approx4$ & \\
&& \textit{Solar min} & $\approx12$ & \\
\enddata
\end{deluxetable}

\section{Observations}\label{sec:obs}
Here, we provide full details of our ultraviolet and X-ray observations for each star (Table~\ref{tab:observations}).

\startlongtable
\begin{deluxetable}{ccccccc}
\tablecolumns{7}
\tablewidth{0pt}
\tablecaption{Observation Summary \label{tab:observations}}
\tablehead{\colhead{Star} &
            \colhead{HST Mode} & 
            \colhead{Date} & 
            \colhead{T$_{exp}$ (s)} &
            \colhead{X-ray mode} & 
            \colhead{Date} &
            \colhead{T$_{exp}$ (s)}
                  }
\startdata	
GJ 4102&STIS G140L&2022-09-29&7377&EPIC*&2020-02-16&19900\\
&STIS G140L&2023-05-24&4525&&2020-09-24&16000\\
&STIS G140M&2023-03-25&7348&&&\\
&STIS G230L&2023-05-24&4786&&&\\
&STIS G430L&2023-05-25&120&&&\\
&&&&&&\\
K2-18&STIS G230L&2022-07-10&6810&\ldots&&\\
&STIS G230L&2022-12-07&2106&&&\\
&STIS G430L&2022-07-10&300&&&\\
&&&&&&\\
GJ 367&COS G130M&2022-01-01&1472&EPIC&2021-12-26&13000\\
&STIS G140M&2022-01-01&2033&&&\\
&STIS G230L&2022-01-01&2057&&&\\
&STIS G430L&2022-01-01&60&&&\\
&&&&&&\\
TOI-776&STIS G140L&2022-05-31&7125&EPIC&2022-12-14&28000\\
&STIS G140L&2022-06-21&4618&&&\\
&STIS G140L&2022-12-20&4618&&&\\
&STIS G140L&2022-12-21&7125&&&\\
&STIS G140M&2022-06-20&4591&&&\\
&STIS G230L&2022-06-20&1991&&&\\
&STIS G230L&2022-06-22&2366&&&\\
&STIS G430L&2022-06-20&120&&&\\
&&&&&&\\
GJ 341&COS G130M&2022-01-20&1592&EPIC&2022-01-23&24700\\
&STIS G140M&2022-01-20&2066&&&\\
&STIS G230L&2022-01-20&2112&&&\\
&STIS G430L&2022-01-20&30&&&\\
&&&&&&\\
TOI-134&COS G130M&2022-04-12&8575&EPIC&2022-04-09&27950\\
&STIS G140M&2022-04-01&7189&&&\\
&STIS G230L&2022-04-01&1998&&&\\
&STIS G430L&2022-04-01&120&&&\\
&&&&&&\\
TOI-260&COS G130M&2023-06-18&4004&EPIC&2023-06-18&18500\\
&STIS G140M&2023-08-21&2022&&&\\
&STIS G230L&2023-08-21&2046&&&\\
&STIS G430L&2023-08-21&60&&&\\
&&&&&&\\
TOI-178&STIS G140L&2022-12-27&2111&HRC&2022-08-25&9270\\
&STIS G140L&2023-06-13&2111&&2023-09-23&14230\\
&STIS G140M&2022-10-25&1989&&2023-09-24&12950\\
&STIS G140M&2023-05-15&2086&&2023-11-03&10860\\
&STIS G230L&2023-06-13&2366&&2023-11-05&9810\\
&STIS G430L&2022-10-25&120&&&\\
&&&&&&\\
NGTS-10&STIS G230L&2022-04-23&16180&\ldots&&\\
&STIS G230L&2022-04-26&2108&&&\\
&STIS G430L&2022-04-23&180&&&\\
&&&&&&\\
TOI-836&STIS G140L&2022-07-28&2058&EPIC&2022-07-25&30500\\
&STIS G140M&2022-07-28&2359&&&\\
&STIS G230L&2022-07-28&2093&&&\\
&STIS G430L&2022-07-28&20&&&\\
&&&&&&\\
K2-141&STIS G140L&2022-10-20&2498&EPIC*&2015-12-09&37900\\
&STIS G140L&2023-05-10&2102&&&\\
&STIS G230L&2022-10-20&2042&&&\\
&STIS G430L&2022-10-20&60&&&\\
&&&&&&\\
HATS-72&STIS G140L&2022-11-07&9640&\ldots&&\\
&STIS G140L&2022-11-09&9640&&&\\
&STIS G230L&2022-12-02&4046&&&\\
&STIS G430L&2022-12-02&180&&&\\
&&&&&&\\
TOI-402&STIS G140L&2022-12-09&1973&HRC&2023-09-12&14270\\
&STIS G230L&2022-12-09&2007&&2023-09-16&14240\\
&STIS G430L&2022-12-09&30&&2023-10-05&14410\\
&&&&&2023-10-06&10890\\
&&&&&2023-10-28&4820\\
&&&&&&\\
&&&&&&\\
TOI-421&STIS G140L&2022-08-28&7086&\ldots&&\\
&STIS G230L&2022-08-28&2101&&&\\
&STIS G430L&2022-08-28&10&&&\\
&STIS G140M*&2022-02-15&1965&&&\\
&&&&&&\\
&&&&&&\\
WASP-63&STIS G230L&2022-04-20&1554&\ldots&&\\
&STIS G430L&2022-04-20&10&&&\\
&&&&&&\\
HD 80606&STIS G140L&2023-10-16&4613&EPIC*&2018-11-15&37900\\
&STIS G230L&2023-10-16&2120&&&\\
&STIS G430L&2023-10-16&5&&&\\
&&&&&&\\
Kepler-51&STIS G230L&2023-01-23&4079&\ldots&&\\
&STIS G430L&2023-01-23&120&&&\\
&&&&&&\\
HIP 67522&STIS G140L&2022-09-03&9605&ACIS*&2021-02-08&2050\\
&STIS G140L&2022-09-05&4587&&&\\
&STIS G230L&2022-09-05&2103&&&\\
&STIS G430L&2022-09-05&10&&&\\
&&&&&&\\
WASP-166&STIS G140L&2022-05-31&7096&HRC&2023-11-01&9920\\
&STIS G140L&2022-06-06&3601&&2023-11-13&8790\\
&STIS G230L&2022-06-06&1754&&2024-02-02&10150\\
&STIS G430L&2022-06-06&10&&&\\
&&&&&&\\
&&&&&&\\
WASP-121&STIS G140L&2022-10-06&6680&EPIC*&2017-04-06&9000\\
&STIS G430L&2022-10-06&10&&&\\
&STIS E230M&2017-02-24&13060&&&\\
&STIS E230M&2017-04-10&13060&&&\\
&STIS G430L&2016-10-24&12210&&&\\
&STIS G430L&2016-11-06&12210&&&\\
&STIS G750L&2016-11-12&11592&&&\\
\enddata
\tablenotetext{*}{Archival data from other programs.}
\end{deluxetable}

\clearpage

\bibliographystyle{aasjournal}
\bibliography{aabib,newcites}



\end{document}